\newif\ifprep\preptrue
\newif\ifprepgood\prepgoodfalse
\ifprep
\documentstyle[twocolumn,aps,epsf,rotate,amsfonts]{revtex}
\else
\documentstyle[aps,preprint,epsf,rotate,amsfonts]{revtex}
\fi
\title{Periodic orbits near bifurcations of codimension two: Classical
mechanics, semiclassics, and Stokes transitions}
\author{Henning Schomerus}
\address{Fachbereich Physik,
	Universit\"at--Gesamthochschule Essen,
			D--45117 Essen, Germany}

\newcommand{\tr}{\mathop{\rm tr}\nolimits}
\begin{document}
\draft
\date{Date: \today}
\maketitle

\begin{abstract}
We investigate classical and semiclassical aspects of codimension--two
bifurcations of periodic orbits in Hamiltonian systems.
A classification of these bifurcations
in autonomous systems
with two degrees of freedom or time--periodic systems with
one degree of freedom is presented.
We derive uniform approximations to be used in semiclassical trace
formulas
and determine also
certain global bifurcations in conjunction with Stokes
transitions that become important in the ensuing diffraction catastrophe
integrals.
\end{abstract}
\vspace{1cm}
\pacs{Pacs: 05.45.+b, 03.65.Sq, 03.20.+i}

\section{Introduction}

Periodic--orbit theory
aims at the semiclassical evaluation of energy
levels of quantum systems and relates their spectral properties
to periodic orbits of the corresponding classical system.
For autonomous systems one considers the trace of the Green's function
$G(E)$ which determines also the density of states $d(E)$.
For periodically driven
systems the object of interest are the traces $\tr F^n$
of the stroboscopic time--evolution operator
over $n$ periods; they encode the so--called quasienergies of
states that are stationary in the stroboscopic description.
Both types of traces can be written as a sum of {\em individual}
contributions of periodic orbits for chaotic
(hyperbolic) systems 
\cite{Gutzwiller:1971,Gutzwiller:1990,Tabor:1983,Junker:Leschke:1992}
or a sum over rational tori for integrable motion
\cite{Berry:Tabor:1976,Berry:Tabor:1977}.

Recent semiclassical studies
\cite{Ozorio:Hannay:1987,Ozorio:1988,Kus:Haake:1993b,Sieber:1996a,%
Schomerus:Sieber:1997a,Sieber:Schomerus:1997a}
based on periodic--orbit theory
were devoted to the neighbourhood, in the space of control parameters,
of classical bifurcations. These
are instances in which
periodic orbits coalesce and are the mechanism how
orbits are born or disappear, or change their configuration when the
energy or an external parameter is varied. 
Bifurcations are ubiquitous in systems with a mixed phase space and
pave the path from integrable to chaotic motion. 

A {\em collective\/} treatment of the bifurcating orbits 
was found necessary, and even more the inclusion of 
predecessors of such orbits which live in complexified phase space and
were termed {\em ghosts\/}.
A collective contribution comes from an orbit cluster and can only
far away from the bifurcation be written as a sum of individual
contributions.
Both types of
contributions are an additive term in the periodic--orbit expansion of
the trace in question.

The existing semiclassical (and most of the classical)
studies focus on the generic
bifurcations in the classification of Meyer and Bruno
\cite{Meyer:1970,Bruno:1970,Bruno:1972} (see also \cite{Ozorio:1988}).
These are the bifurcations that are
typically encountered when one has only a single parameter at hand
to steer the system through parameter space, or, equivalently,
when one investigates the
periodic--orbit families in a given autonomous system as a function of energy.
In general one assigns a codimension to each type of bifurcation by
counting the number of parameters to be controlled in order to encounter it
in a general setting.
(The class of bifurcations of a given codimension is enlarged when
symmetries are imposed on the system
\cite{Rimmer:1983,Golubitsky:Stewart:1987,Aguiar:Malta:1987,Aguiar:Malta:1988%
,Ozorio:Aguiar:1990}.) 
The generic bifurcations are accordingly the bifurcations of codimension one.
In each of these bifurcations
there is a central orbit of period $n$, surrounded 
by one or two satellite orbits of period $nm$. The cases $m=1,2,3,4$ are
the tangent, period--doubling, period--tripling, and period--quadrupling
bifurcations, respectively. There are two types of
period--quadrupling bifurcations (island chain and touch--and--go),
but only one for all other $m$.
All period--$m$ bifurcations with $m\geq 5$ follow the island chain pattern.

In \cite{Ozorio:Hannay:1987,Ozorio:1988,Kus:Haake:1993b} {\em transitional\/}
approximations for the collective contributions 
were derived that are only valid close to
the (generic) bifurcation; far away  from the bifurcation they give rise to
individual contributions with the wrong amplitudes.
Refs. \cite{Sieber:1996a,Schomerus:Sieber:1997a,Sieber:Schomerus:1997a}
give {\em uniform\/} approximations that are
valid even far away from the bifurcation, where they
asymptotically
split into individual contributions with the
correct amplitudes.

The present work is devoted to bifurcations of codimension two
in autonomous Hamiltonian systems 
with two degrees of freedom or time--periodic Hamiltonian systems with  
one degree of freedom. The bifurcations are classified and their impact
on
semiclassical periodic--orbit theory is studied in detail. We derive
uniform
approximations of collective contributions to the
semiclassical traces and discuss
certain global
bifurcations in conjunction with so-called Stokes transition.

Bifurcations of codimension two manifest themselves in one--parameter studies in
certain sequences of generic bifurcations.
Sadovski{\'\i}, Shaw, and Delos
\cite{Sadovskii:Shaw:1995,Sadovskii:Delos:1996}
found that such sequences can be explained
by normal--form theory
\cite{Poincare:1957,Birkhoff:1927,Deprit:1969,Arnold:1988},
but did not attempt a classification with respect to the codimension.
The classical part of the present study is very much inspired by
these works.

It was indeed demonstrated in Ref.\ \cite{Schomerus:1997a}
that bifurcations of codimension two
are frequently felt semiclassically and necessitate a collective treatment
even when one steers the system through control
space with less than two parameters.
This implies that 
collective contributions of this kind will constitute a basic ingredient
in a semiclassical trace formula for systems with a mixed phase space.
They indeed played an essential r{\^ole} in the semiclassical
determination of the quasienergies of the kicked top
\cite{Schomerus:Haake:1997}.

The paper is organized as follows.
In section \ref{sec:nform} we derive
normal forms for the Hamiltonian that describe the bifurcations of
codimension two. We find that they are organized by the multiplicity
$m$ in analogy to the situation for codimension one.
The corresponding  sequences of codimension one bifurcations 
involve a tangent
bifurcation of period $nm$, followed by a period--$m$ bifurcation that
involves another orbit of period $n$.
(The representative case
studied in \cite{Schomerus:1997a} corresponds to $m=3$.)

The normal forms and the sequences of
codimension one bifurcations in the neighbourhood of the codimension 
two point in control space (technically spoken, the {\em unfolding} of
the normal forms)
are discussed and illustrated in section \ref{sec:discn}.

With section \ref{sec:deriv} we turn to aspects within semiclassical
periodic--orbit
theory and present the
starting point for the derivation of collective contributions of
bifurcating orbits, consisting of two--dimensional
integrals over phase space or
Poincar{\'e} surfaces of section that involve the generating function
$\hat S$ of the classical stroboscopic map.

Uniform approximations of these contributions are
derived in section
\ref{sec:unif}. They involve a phase
function $\Phi$ and an amplitude function $\Psi$. Normal forms of these
are obtained from the corresponding Hamiltonians by non--canonical
transformations and partial integrations. Investigating the influence
from higher--order terms in the
phase equips us with a sufficient number of coefficients to guarantee
the right stationary--phase limit
of the expressions, which are therefore truly uniform.
[The cases $m=1,2$ lead to 
standard diffraction
integrals connected to the cusp and butterfly catastrophes,
respectively. Among the large number of applications, the transitional
approximations have been investigated in connection to
bifurcations of closed orbits in \cite{Main:Wunner:1997}.
Uniform approximations have not
been derived there, however, and the canonical invariant determination of
coefficients as well as Stokes transitions are also not discussed.]

In section \ref{sec:stokes} we discuss certain global bifurcations
that become important in the ensuing diffraction catastrophe integrals.
They give rise to 
{\em Stokes transitions}
in which the contribution of a
ghost satellite is switched on or off.
The ghosts and transitions arise when the integrals are analyzed
using the method of steepest descent.
Stokes transitions have been investigated in the context of diffraction
integrals and asymptotic expansions before.
A uniform approximation for an isolated
transition is given in \cite{Berry:1989}
and has been applied for
perturbed cat maps in \cite{Boasman:Keating:1995},
which is the only treatment of this
phenomenon in semiclassics that we know of. The
Stokes transitions investigated there, however, 
occur far away from any other bifurcation and can
be regarded as isolated.
A transition requires special treatment
when it occurs in the immediate neighbourhood of a usual `local'
bifurcation, a situation that is often encountered in mixed systems.
The uniform approximations and
normal forms derived here
can also be employed to describe
the Stokes transition of a period--$nm$ ghost prior to a tangent
bifurcation when the so--called `dominant' orbit involved is real and of
period $n$.
The complete sequence of local and
global bifurcations that we can handle consists of the period--$m$
bifurcation at the central orbit and tangent bifurcations of satellites,
followed by Stokes transitions in which ghost satellites once more
interact with the central orbit.

We conclude and point out open questions in section \ref{sec:disc}.

\section{Normal forms of the {Hamiltonian}
for bifurcations of codimension two}
\label{sec:nform}
\subsection{Objective}

The local bifurcations to be discussed are instances in which periodic orbits
coalesce as parameters are varied.
The types of bifurcations generically encountered in a given class of
systems depends on the number of parameters
varied, and the number of parameters typically
needed to be controlled in order to
find a particular type is called its codimension.
Here we investigate bifurcations of codimension two in the class of
periodically time--dependent Hamiltonian systems with one degree of freedom.
In other words, we study families of Hamiltonians 
\begin{equation}
\label{eq:hamfam}
H(q,p,t;\varepsilon,a)
\end{equation}
that depend on the two parameters $\varepsilon$ and $a$ and obey
$H(t)=H(t+1)$, where the period is set to unity for convenience.
In general these systems have no time--reversal nor any geometric symmetry.
The discussion directly carries over to autonomous systems with two degrees of
freedom since these can be reduced
to one--parameter
families of periodic systems with one degree of freedom
by a standard procedure described e.\,g.\ in \cite{Ozorio:1988}.

\subsection{The bifurcation condition}
\label{sec:bifcond}

The periodic orbits show up as fixed points in iterations of the so--called
stroboscopic map $(q,p)\to(q',p')$
which is induced by the evolution over one period.
This map is area preserving. Its linearized version
\begin{equation}
M=\left(
\begin{array}{cc}
\displaystyle
\left.\frac{\partial q'}{\partial q}\right|_{p}
&
\displaystyle
\left.\frac{\partial q'}{\partial p}\right|_{q}
\\
{}
\\
\displaystyle
\left.\frac{\partial p'}{\partial q}\right|_{p}
&
\displaystyle
\left.\frac{\partial p'}{\partial p}\right|_{q}
\end{array}
\right)
\label{eq:linmap}
\end{equation}
(corresponding to a
$2\times 2$ matrix) hence obeys $\mbox{det}\, M=1$. 
Orbits that appear for the first time in the $n_0$th iteration are said to
be of primitive period $n_0$. Such an orbit gives rise to $n_0$ fixed
points in each $n$--step map with $n=rn_0$, where $r$ is an integer counting
repetitions.

The eigenvalues $\lambda_{1,2}$ of the linearized  $n_0$--step map
$M^{(n_0)}$ are
reciprocal to each other. A stable orbit has unimodular
eigenvalues and hence $\tr M^{(n_0)}=2\cos \omega$ with the real stability
angle $\omega$. An orbit is instable if the eigenvalues are real. There
are two cases depending on the sign of the eigenvalues, $\tr M^{(n_0)} = \pm
2\cosh \omega'$, with the real and by convention positive instability
exponent $\omega'$.

In general,
an orbit bifurcates whenever the linearized $n$--step map $M^{(n)}$ (with
again $n=rn_0$) acts at the locus of the orbit in phase space
in at least one direction as the identity map and hence obeys
\begin{equation}
\tr M^{(rn_0)}=
\tr \left(M^{(n_0)}\right)^r=
2
\;,
\end{equation}
or, equivalently, 
\begin{equation}
\label{eq:bifcond}
\tr M^{n_0}=2\cos(2\pi l/m)
\;,
\end{equation}
where the integers $l$, $m$ are taken as relatively prime.
This bifurcation condition implies a discrete $m$--fold
rotational symmetry $C_m$ in
the flow pattern around the
bifurcating orbit.
For $m\geq 2$ the orbit in question
is a `central' orbit on which `satellite'
orbits of primitive period $n_0 m$ contract at the bifurcation.
For $m=1$ there are two possibilities, the orbit is either involved as a
satellite in a bifurcation with an orbit of smaller primitive period, or
it takes part in an isochronous bifurcation with other orbits of same
period.
Turning these observations around, there is always
a central periodic orbit of
smallest primitive period $n_0$ among the bifurcating orbits which
coalesces with satellites of period $n_0m$.
For that reason $m$ is called the multiplicity.

The bifurcations of codimension one have been classified by Meyer and
Bruno \cite{Meyer:1970,Bruno:1970,Bruno:1972} (see the introductory
section).
They constitute the building blocks of the scenarios of higher codimension
and will be illustrated together with those of codimension two
in the next section.
Recall
that for each $m$ there is exactly one type
with the exception of $m=4$ which allows for two variants.

\subsection{Classification of normal forms}

The bifurcation condition (\ref{eq:bifcond})
is reflected by the Hamiltonian
flow around the central orbit;
accordingly, the bifurcations can be investigated
by studying the general form of the 
Hamiltonian in the vicinity of this orbit.
Following Refs.
\cite{Ozorio:Hannay:1987,Ozorio:1988,Sadovskii:Shaw:1995,Sadovskii:Delos:1996,%
Meyer:Hall:1992},
we aim at the reduction of the general expressions
to certain simple normal forms by suitable canonical transformations.
In that way one can identify the parameters that govern the
distance to the bifurcation.
They can be chosen such that
$\varepsilon=0$ brings us on a codimension one bifurcation and the 
codimension two scenario is encountered if in addition $a=0$. 

For codimension one the construction
(that is carried out for that case in detail in \cite{Ozorio:1988} and
is recapitulated below)
leads to the Birkhoff normal forms
\begin{equation}
\label{eq:hnform}
\begin{array}{|c|l|}
\hline
m & h^{(m)}(q,p)-H_0 \\
\hline
1 & \varepsilon q + a q^3 + \frac \sigma 2 p^2\\
2 & \varepsilon q^2 + a q^4 + \frac \sigma 2 p^2\\
3 & \varepsilon I + a I^{3/2}\cos 3\phi + b I^2 \\
4 & \varepsilon I + a I^2(1+\cos 4\phi) + b I^2(1-\cos 4\phi)
\\
\geq 5 &   \varepsilon I + a I^2 + \sum_{l=3}^{[m/2]}b_lI^l
+c I^{m/2}\cos m\phi\\
\hline
\end{array}
\;.
\end{equation}
For $m \geq 3$ they have been expressed in 
canonical polar coordinates $I$, $\phi$  with
\begin{equation}
\label{eq:canonpolarcoor}
q = \sqrt{2 I} \sin \phi \;, \qquad p= \sqrt{2 I} \cos \phi \;.
\end{equation}
The quantity $H_0$ is a constant.

These expressions are autonomous and
display the $m$--fold symmetry even globally.
The periodic orbits are mapped to
fixed points $\frac {\partial H}{\partial q}=0$,
$\frac {\partial H}{\partial p}=0$ and are thus determined as roots
of polynomials in $p$ and $q$.
From
Vieta's relations between these roots (or locations of satellites)
and the coefficients of the polynomials
it follows that orbits
collapse on the center as the lowest--order terms $\sim \varepsilon$ (and
$\sim a$ for codimension two) of the fixed--point
equations are steered to zero.

To describe the codimension two variants 
we have to include higher--order terms to account for additional satellites
that approach the center 
and obtain the extended normal forms
\begin{equation}
\label{eq:Hnform}
\begin{array}{|c|l|}
\hline
m & H^{(m)}(q,p)-H_0 \\
\hline
1 & \varepsilon q + a q^3 + b q^4 + \frac \sigma 2 p^2\\
2 & \varepsilon q^2 + a q^4 + b q^6 + \frac \sigma 2 p^2\\
3 & \varepsilon I + a I^{3/2}\cos 3\phi + b I^2 \\
4 & \varepsilon I + a I^2(1+\cos 4\phi) + b I^2(1-\cos 4\phi)\\
& \qquad + cI^3 (1+\cos 4\phi)
\\
5 &   \varepsilon I + a I^2 + b I^{5/2}\cos 5\phi\\
6 &   \varepsilon I + a I^2 + b I^3 + c I^3\cos 6\phi \\
\geq 7 &   \varepsilon I + a I^2 + \sum_{l=3}^{[m/2]}b_l I^l
+ c I^{m/2}\cos m\phi 
\\
\hline
\end{array}
\;.
\end{equation}
The normal forms for
$m\geq 5$ are the usual Birkhoff normal forms; they will, however,
be investigated not only for small $\varepsilon$ but also for small $a$.
The expressions for $m\leq 4$ go exactly one order beyond the Birkhoff
normal forms.

\subsection{Derivation of normal forms}

In the derivation of the normal forms (\ref{eq:hnform}) and (\ref{eq:Hnform})
the central orbit is placed into the origin of a local coordinate
system $(q,p)$ by a time--dependent 
canonical transformation.
This is done in such a way that a Taylor expansion in $q$ and $p$ yields 
\begin{equation}
\label{eq:hstart}
H(q,p,t)= H_0+\frac \omega 2 q^2+\frac \sigma 2 p^2+{\cal O}(3)
\end{equation}
with time--independent $\omega$ and $\sigma$. 
The remainder in the expansion
indicates third and higher orders in $q$ and $p$.
The new Hamiltonian will be $n_0$--periodic but otherwise as general as
the one we started with. Hence it suffices to study the case $n_0=1$. 

One passes to a rotating coordinate system,
where the angular frequency $2\pi l/m$
is adopted to the motion around the center at the bifurcation, and examines
the expansion of the Hamiltonian in a Taylor series in $q$
and $p$ as well as a Fourier series in $t$.  Most terms in
the expansion can be removed by canonical transformations (specified below)
up to a certain order in $p$, $q$ and the deviations of the
parameters $\varepsilon$, $a$ that govern the distance to the bifurcation.
The expansion is carried out up to a certain degree,
requiring that no bifurcating orbits are added or
qualitatively affected by the omitted higher--order terms.

\subsubsection{The cases $m=1,2$}

The linearized map $M$ has degenerate eigenvalues $1$ for $m=1$ and $-1$
for $m=2$. This entails that
the corresponding bifurcation scenarios of codimension one and two
are essentially one--dimensional. The reason is that
the linearized map is
in these cases generically not diagonalizable:
Otherwise one would have
located the matrix $M=\pm 1$
in the three--dimensional manifold $Sp(2,{\Bbb
R})$
of real
$2\times 2$ matrices with $\mbox{det}\,M=1$ \cite{Meyer:Hall:1992}.
With only two parameters at
one's disposal, however, one
generically finds only such matrices with $\tr M=\pm 2$
for which the eigenspace is one--dimensional.
The linearized map describes then a shear transformation
and acts as the identity in only one direction.

The reasoning can be put onto another footing by resolving the apparent
paradox that we can cast the second--order terms into a diagonal form
$\omega q^2/2 +\sigma p^2/2$ in the immediate neighbourhood of the
bifurcation, which involves only two parameters.
This diagonalization,
however,
depends  in a singular way on the parameters of the original expansion
$a_1 q^2 + a_2 p^2 + a_3 qp$. (Note that
all three terms are $C_1$-- and $C_2$--invariant.)
Accordingly, three coefficients $a_i$ are to be controlled
in order to let the second order vanish.
Such an argumentation will be used again for $m=4$.

We choose the
$q$--axis as the line on
which orbits approach each other at the bifurcation
and invoke the splitting lemma
of catastrophe theory
in order to simplify the expression in $p$--direction.
With a suitable scaling transformation one can achieve
$|\sigma|=1$.
The normal form for $m=2$ is
symmetric in $q$ since in that case the coordinate 
system rotates by $\pi$ in each period.

\subsubsection{The cases $m\geq 3$}

For $m\geq 3$ the central orbit is stable close to the bifurcation
so that we can achieve $\sigma=\omega$ in
(\ref{eq:hstart}), entailing $\tr M=2\cos\omega$. 
It is then convenient to use 
canonical polar coordinates
(\ref{eq:canonpolarcoor}).
In the rotating coordinate system
the leading--order term in $I$ takes the form $\varepsilon I$ with
$\varepsilon=\omega-2\pi \frac l m$ in agreement with the bifurcation
condition (\ref{eq:bifcond}).
The expansion reads in detail
\begin{eqnarray}
H&=&H_0+\varepsilon I + \sum_{k=3}^{\infty}\sum_{l'=-\infty}^\infty\sum_{m'}
V_{kl'm'}I^{k/2}
\\
\nonumber
&&\times
\cos\left[m'\phi+2\pi\left(m'\frac lm-l'\right)t+\phi_{klm}\right]
\;,
\end{eqnarray}
where $m'$ runs from $-k,-k+2,\ldots,k$  since only such terms arise
from expressions of type $q^{k'}p^{k-k'}$.
Let us assume that all $t$ and 
$\phi$ dependence is already eliminated up to a
certain order $I^{k/2}$.
The majority of
terms of this order are then removed by a canonical
transformation to new coordinates $J,\theta$ that is generated
according to $\frac{\partial G}{\partial I}=J$, $\frac{\partial
G}{\partial \theta}=\phi$, $H'=H-\frac{\partial G}{\partial t}$ by the
function
\begin{eqnarray}
G(\theta,I)&=&I\theta
+I^{k/2}\sum_{l'=-\infty}^\infty\sum_{m'}
G_{kl'm'}
\\
\nonumber
&&
\times
\sin\left[m'\theta+2\pi\left(\frac {m'l}m-l'\right)t+\phi_{klm}\right]
\end{eqnarray}
with 
\begin{equation}
G_{kl'm'}=\frac{V_{kl'm'}}{\omega m'-2\pi l' }
=\frac{V_{kl'm'}}{\varepsilon m'+ 2\pi lm'/m-2\pi l' }
\;.
\end{equation}
After the transformation
we switch back in our notation from $J,\theta$ to $I,\phi$. 
The coefficient $G_{kl'm'}$
diverges at the bifurcation if the resonance condition
$l'=l\frac {m'} m$ is met. This affects, for instance,
all $t$-- and $\phi$--independent terms ($l'=m'=0$).
The remaining $\phi$--dependent terms are of type $I^{k/2}\cos(nm \phi - \tilde
\phi_{kn})$. Here $n=1,2,3,\ldots$ is an integer since $l$ and $m$ are
relatively prime, 
and $k\geq nm$ as before.
In the orders that appear in the
normal forms the latter restriction admits only $n=1$.
The $\phi$--dependent term
of lowest order in $I$ is generically
of type $I^{m/2}\cos\left(m\phi+\tilde \phi_{m1}\right)$.
A shift of $\phi$ eliminates the constant in the cosine. If the
coefficient of this term is not small then one can get rid of
constants $\tilde \phi_{k1}$ in higher orders $k > m$
by a transformation of the form
$\phi=\theta+\sum_{k'=1}^\infty g_{k'} I^{k'}$ with suitably chosen
coefficients.

\subsubsection{Further reduction for $m=4$}

Additional considerations are needed for $m=4$. The most general
expression that goes one order beyond the Birkhoff normal form reads
\begin{eqnarray}
\nonumber
\tilde H^{(4)}&=&H_0+\varepsilon I + aI^2(1+\cos4\phi) + bI^2(1-\cos4\phi)
\\ &&
+ (c+d)I^3+(c-d)I^3\cos4\phi+ eI^3\sin4\phi
\label{eq:h4tilde}
\;.
\end{eqnarray}
The codimension two bifurcation is approached for vanishing 
$\varepsilon$ and $a$ (or $b$), while the other second--order coefficient
$b$ (or $a$) is finite.
Both cases are
equivalent and mapped onto each other by a rotation about $\pi/4$. The
normal form $H^{(4)}$ has been written down for small $a$. Since $b$ is
finite we can eliminate two of the three third--order
terms in (\ref{eq:h4tilde}) by a canonical transformation and achieve $d=e=0$.
[The corresponding generating
function is of the simple form
$G=\theta I-(dI^2\sin 4\theta)/(8b)-eI^2/(8b)$
if corrections involving
$\varepsilon$ and $a$ are neglected; the complete form is slightly
more complicated.]

The fine--print in the derivation for $m=4$ is that two orbits pretend to
bifurcate also as we send $a\to b$. We identify now these orbits
and show that for codimension two they are
actually ghosts (complex solutions of the fixed--point equations)
at a finite distance to the center.
For simplicity we set $e=0$; this does not affect the general line of
reasoning.
The satellites that concern us solve
the fixed--point equation
\begin{equation}
\frac{\partial \tilde H^{(4)}}{\partial \phi}= -4I^2[a-b+(c-d)I]\sin 4\phi=0
\end{equation}
by $I=I^{(0)}\equiv (b-a)/(c-d)$. The other fixed--point
equation yields
\begin{equation}
\label{eq:cosm4}
\cos 4\phi=-3\frac{c+d}{c-d}-\varepsilon
\frac{c-d}{(a-b)^2}+2\frac{a+b}{a-b}\equiv C^{(0)}
\;.
\end{equation}
(These satellites
are related by the reflection symmetry $\phi\to -\phi$  and undergo a
pitchfork bifurcation as $|C^{(0)}|=1$. The symmetry is broken if
$e\neq 0$.)
For reasons similar to those put forward for $m=1,2$, {\em two}
parameters have to be controlled in order achieve $I^{(0)}=0$, i.\,e.,
$a=b$:
For the given
parameter combination there is no $\phi$ dependence in the
second order of $I$.
But actually there are two independent terms involving $\phi$, namely
$\cos 4\phi$
and $\sin 4\phi$ --- one of them had been eliminated by a diagonalization
that is again sensible only if the other
has a non--vanishing coefficient. We already used two
parameters, then, such that we 
must assume
$\varepsilon$ and $a\approx b$ to be finite.
This gives $|\cos 4\phi|\sim(a-b)^{-2}\gg 1$ and $\cos
\phi \sim (a-b)^{-1/2}={\cal O}({I^{(0)}}^{-1/2})$,
and the Cartesian coordinates
(\ref{eq:canonpolarcoor}) indeed remain finite:
As announced this shows that
the orbits that appeared to
bifurcate are complex solutions (with real $p$ and imaginary $q$)
of the fixed--point equations and stay away from the center.

\section{Local bifurcation scenarios}
\label{sec:discn}

We discuss now in detail 
the bifurcations of codimension two that are described by the
normal forms given in the preceding section.
In each case the location of the periodic points, given as the
solutions of the
fixed--point equations 
\begin{equation}
\frac{\partial H}{\partial q}=0\;,\qquad
\frac{\partial H}{\partial p}=0\;,
\end{equation}
are investigated as the parameters are varied.
Sequences of codimension one bifurcations 
are encountered if only one parameter is varied close to a
codimension two point  
\cite{Sadovskii:Shaw:1995,Sadovskii:Delos:1996}. They are discussed here for
fixed $a$ and variable $\varepsilon$ and
are illustrated by contour plots of the normal forms.
Unstable orbits appear there as saddles while stable orbits correspond to
maxima or minima.
In all these sequences there is a period--$m$ bifurcation at
$\varepsilon=0$ and a tangent bifurcation of satellites at the 
parameter combinations
\begin{equation}
\begin{array}{|c|l|}
\hline
 m & \mbox{tangent bifurcation of satellites}\\
\hline
 1 & \varepsilon=-\frac 14 \frac{a^3}{b^2}\\
 2 & \varepsilon=\frac 13 \frac{a^2}{b}\\
 3 & \varepsilon=\frac 9 {32} \frac{a^2}{b}\\
 4 & \varepsilon=\frac 2 {3} \frac{a^2}{c}\\
 5 & \varepsilon=-\frac {128} {675} \frac{a^3}{b^2}\\
 6 & \varepsilon=-\frac 1 3  \frac{a^2}{b \pm c}\\
 \geq 7 & \varepsilon\approx \frac 13 \frac{a^2}b\\
\hline
\end{array}
\;.
\end{equation}
Global bifurcations that are of particular interest
in the context of uniform approximations are discussed in section
\ref{sec:stokes}.

\subsection{Tangent bifurcations ($m=1$)}

In a tangent bifurcation two orbits of the same primitive period
coalesce. On
one side of the bifurcation both orbits are ghosts,
i.\,e., complex solutions of the
fixed--point equations, and their
coordinates and other characteristic quantities 
are related by complex conjugation.
On the other side of the bifurcation
both orbits are real, one of them being initially
stable and the other unstable.
The scenario is described by the
normal form  $h^{(1)}$ which accounts for two periodic orbits $\pm$ at
coordinates $p_\pm=0$ and
\begin{equation}
q_\pm=\pm\sqrt{-\frac 13 \frac{\varepsilon}{a}}\;.
\end{equation}
Often one encounters a third orbit of identical period
in close neighbourhood (in phase space)
to the bifurcating orbits. This orbit must be taken into account, for
instance, to obtain a reasonable semiclassical approximation.
One has to work then with the extended normal form $H^{(1)}$.
The fixed--point equation $\partial H/\partial q=0$
is a real cubic polynomial in $q$ and has
three solutions. The number of real solutions is determined by the sign
of the discriminant
\begin{equation}
D=\left(\frac 18 \frac{\varepsilon}{ b}\right)^2
+\frac 14 \frac {\varepsilon}{ b}
\left(\frac 14 \frac { a}{ b}\right)^3\;.
\end{equation}
There are three real solutions for $D<0$ and only one for $D>0$ which
is then accompanied by two complex ones.
Tangent bifurcations are encountered at $D=0$, that is $\varepsilon=0$
or $\varepsilon=- a^3/(4 b^2)$.

\begin{figure}
\ifprep
\epsfxsize16cm
\epsfbox{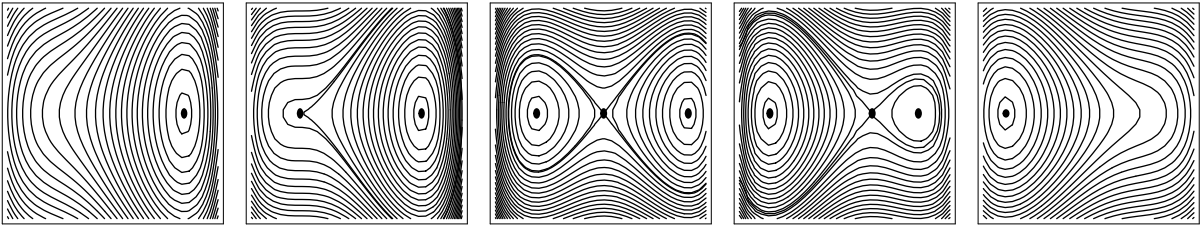}
\fi
\caption{Contour plots of the normal form $H^{(1)}$ as the parameters are
steered to cross two tangent bifurcations close to the bifurcation point of codimension
two.
Initially, only one orbit is present. Two new orbits are born in a
tangent bifurcation. One of them approaches the first orbit, and both
annihilate in an inverse tangent bifurcation.
A similar scenario exists in which stable orbits
are unstable and vice versa; it is obtained by reversing the sign of
$\sigma$.
}
\label{fig:m1cnt}
\end{figure}

A sequence of these two tangent bifurcations is depicted in Figure
\ref{fig:m1cnt}. The codimension two bifurcation 
is obtained when $\varepsilon$ and $a$ pass zero simultaneously. If this
is done in such a way that the discriminant changes sign then
 the number of solutions changes from
one to three in a pitchfork bifurcation. Such a bifurcation is even of
codimension one if the system is time--reversal symmetric or has
a reflection symmetry
\cite{Rimmer:1983,Golubitsky:Stewart:1987,Aguiar:Malta:1987,Aguiar:Malta:1988%
,Ozorio:Aguiar:1990}.

\subsection{Period--doubling bifurcation ($m=2$)
and tangent bifurcation of satellites} 

\begin{figure}
\ifprep
\epsfxsize16cm
\epsfbox{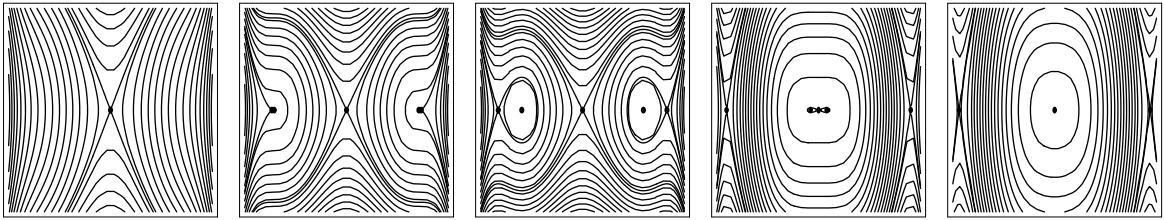}
\fi
\caption{
The typical
sequence for $m=2$ of a tangent bifurcation of period--two satellites and a period--doubling
bifurcation is illustrated by contour plots of $H^{(2)}$.
As for $m=1$ there
exists a similar scenario for the opposite sign of $\sigma$
in which the stability of orbits is changed. The tangent bifurcation would not be encountered
in real phase space 
if the satellites meet at a negative value of $I$.
}
\label{fig:m2cnt}
\end{figure}

In a period--doubling bifurcation the central orbit changes its
stability by absorbing or emitting a satellite of double period.
In the Birkhoff normal form $h^{(2)}$
the central orbit sits at coordinates $q_0=p_0=0$ and the satellite
is represented by two fixed points
\begin{equation}
p_1=0\;, \quad q_1=\pm\sqrt{-\frac{\varepsilon}{2a}}
\;.
\end{equation}
In the extended normal form $H^{(2)}$
the central orbit lies again at $q_0=p_0=0$, but
there are now  two satellites $\pm$ with coordinates
\begin{equation}
\label{eq:m2pos}
q_\pm^2=-\frac 13 \frac ab\pm\sqrt{\frac 19 \frac{a^2}{b^2}-\frac 13
\frac \varepsilon b}
\;.
\end{equation}
A tangent bifurcation of the satellites is encountered at 
\begin{equation}
\varepsilon=\frac 13 \frac{a^2}{b}
\;,
\end{equation}
but the condition $ab<0$ must be obeyed since otherwise both orbits are
still ghosts with
purely imaginary $q$--coordinates.
A sequence of tangent bifurcation of the satellites and 
period--doubling bifurcation (with $ab<0$) is shown in Figure \ref{fig:m2cnt}.

\subsection{Period--tripling bifurcation ($m=3$) and tangent bifurcation
of satellites} 

\begin{figure}
\ifprep
\epsfxsize16cm
\ifprepgood
\epsfbox{m13.ps}
\else
\vspace{-.5cm}
\epsfbox{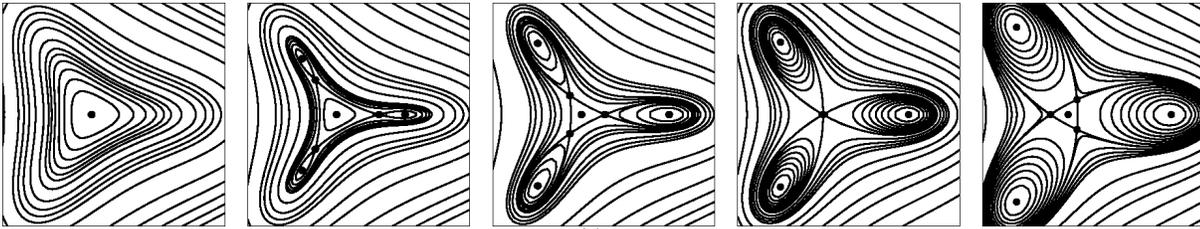}
\fi
\fi
\caption{The contour plots of the normal form $H^{(3)}$ display a
sequence of a tangent bifurcation of satellites and a period--tripling
bifurcation.
}
\label{fig:m3cnt}
\end{figure}

The situation for the period tripling
is visualized in the sequence of contour plots
in Figure \ref{fig:m3cnt}.
Initially, 
a stable periodic orbit of period one
is surrounded by its stability
island. At a certain value of the control parameter
two satellites of triple period come into existence via a tangent
bifurcation. Then the inner (unstable) satellite approaches the
central orbit, collides with it in the period tripling,
and finally re--emerges on the other side.
This scenario has been investigated, for instance, 
in the diamagnetic Kepler problem \cite{Mao:Delos:1992} and for
the kicked top \cite{Schomerus:1997a}.

The Birkhoff normal form $h^{(3)}$ describes the central orbit at $I=0$
and the unstable satellite that is involved in the tripling.
The $\phi$--coordinate of the satellite obeys
$
\partial h^{(3)}/\partial \phi = -3a I^{3/2} \sin 3\phi =0
$.
Since a threefold symmetry is implied by this equation 
it suffices 
to consider the second equation $\partial h^{(3)}/\partial I=0$
on the $p$--axis
after switching back to the coordinates $p,q$, yielding
$ \varepsilon p+\frac 3 {\sqrt 8} a p^2=0 $. Hence the canonical radial
coordinate of the
satellite is
$I=p^2/2=4\varepsilon^2/(9a^2)$.

In the extended normal form $H^{(3)}$ the 
$\phi$--coordinates of the satellites once more obey
$ -3a I^{3/2} \sin 3\phi =0 $.
On the $p$--axis they satisfy now
$ \varepsilon p+\frac 3{\sqrt 8} a p^2 + b p^3=0 $.
This equation has three solutions,
\[
p_0=0\;, \quad p_{\pm}=-\frac {3}{4\sqrt2} \frac a b \pm \sqrt{\frac 9
{32} \frac {a^2}{b^2}-\frac {\varepsilon}{b}}\;.
\]
One in fact sees that the inclusion of the next--order term 
implies the existence of a further satellite. 
At $\varepsilon=9a^2/(32b)$ the satellites undergo a tangent
bifurcation and for $\varepsilon/b>9a^2/(32b^2)$ both satellites are
ghosts.
For $0<\varepsilon/b<9a^2/(32b^2)$ both satellites are on the same side of the 
central orbit,
 while after the period tripling ($\varepsilon=0$)
they lie opposite to each other.
In the limit $\varepsilon /b\to-\infty$
the satellites form a broken torus, well separated from the central orbit.
When a second parameter is varied to
achieve $a=\varepsilon=0$, both satellites are contracted onto the central orbit
in the codimension two bifurcation. 

\subsection{Period--quadrupling bifurcation ($m=4$) and tangent bifurcations
of satellites} 

There are two variants of the period--quadrupling bifurcation depending
on the magnitude of the coefficients $a$ and $b$ in the normal form
$h^{(4)}$. In both cases there are two satellites of quadruple period
involved that lie at
$\sin 4\phi=0$ and are distinguished by the quantity $\cos 4\phi=\pm 1
\equiv\sigma$.
Their radial distance is given by
$I^{(\sigma=1)}= -\varepsilon/(4a)$
and  $I^{(\sigma=-1)}=-\varepsilon/(4b)$.
In the
touch--and--go case $\mbox{sign}\,a=-\mbox{sign}\,b$
an unstable satellite 
becomes a ghost while in turn a ghost solution becomes
real and emerges from the central orbit. In the island--chain scenario
$\mbox{sign}\,a=\mbox{sign}\,b$
there are two ghost  satellites on one side of the bifurcation and two
real satellites on the other, one of them being stable and the other
unstable. 

\begin{figure}
\ifprep
\epsfxsize16cm
\epsfbox{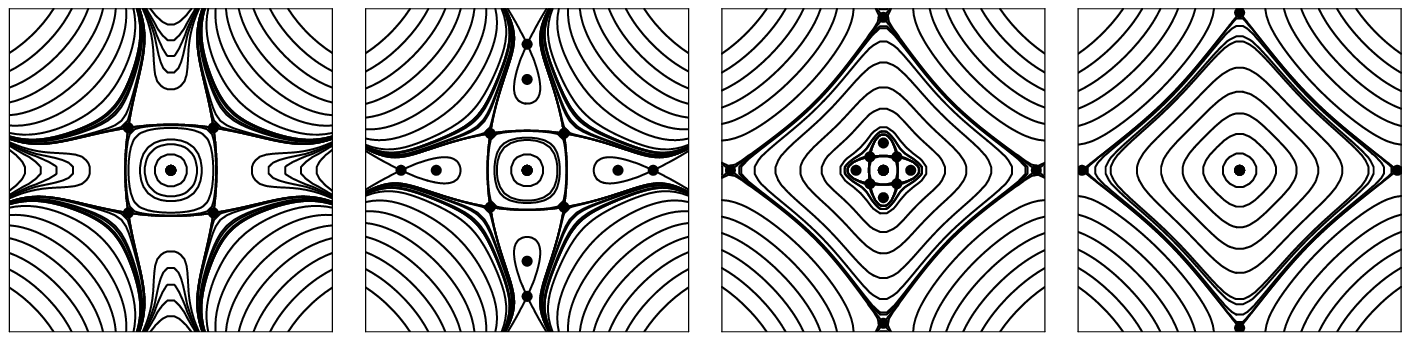}
\epsfxsize16cm
\epsfbox{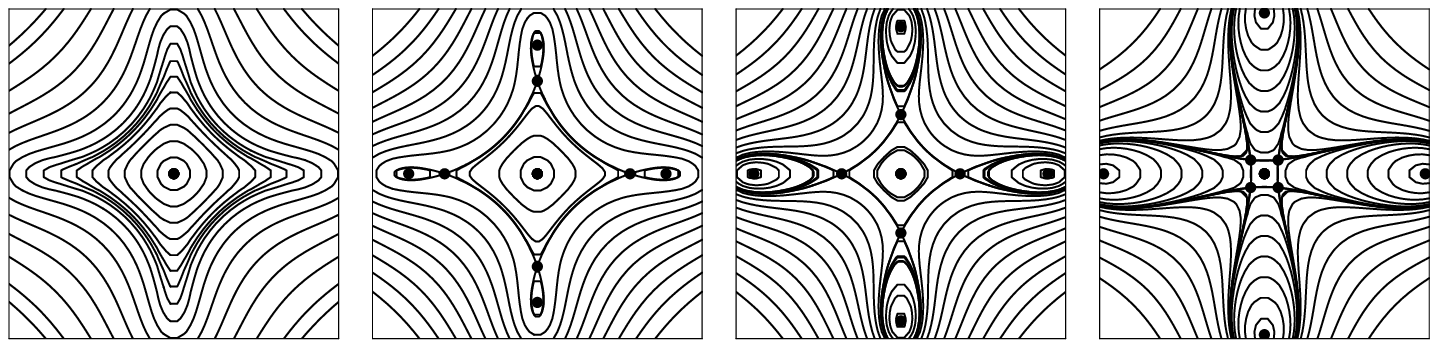}
\fi
\caption{The two sequences of contour plots of $H^{(4)}$ display a
tangent bifurcation of satellites followed by a period--quadrupling
bifurcation. The latter is encountered in its
island--chain version above; below we
have the touch--and--go scenario.
}
\label{fig:m4cnt}
\end{figure}

The next--order terms in the extended normal form $\tilde H^{(4)}$
[eq.\ (\ref{eq:h4tilde})]
involve three new parameters  $c$, $d$ and $e$ and
give rise to six satellites. For $e=0$ there are two satellites on
the lines $\cos 4\phi=1$ and two on the lines $\cos 4\phi=-1$ as well as
the two satellites discussed in the
derivation of the normal form $H^{(4)}$.
There is a tangent bifurcations
at $2a^2=3\varepsilon c$ where the satellites on $\cos 4\phi=1$
coalesce, and another one
at $2b^2=3\varepsilon d$ that involves the satellites on $\cos 4\phi=-1$.
A great variety of possible configurations of all six satellites exists. 
Here
we are, however, only concerned with the codimension two bifurcation, 
described by $H^{(4)}$ and encountered for
$\varepsilon=a=0$. It involves only three 
satellites,
those on the lines $\cos 4\phi=1$
with radial coordinates
\begin{equation}
I^{(1)}_\pm=
-\frac 13 \frac{a}{c}\pm\sqrt{
\frac 19 \frac{a^2}{c^2}-\frac 16\frac\varepsilon{c}}
\label{eq:m4i1}
\end{equation}
and that
satellite with  $\cos 4\phi=-1$ which is closer to the center and lies
with $H^{(4)}$
at $I^{(-1)}=-\varepsilon/(4b)$.
Compared to the situation described by $\tilde H^{(4)}$
the second satellite on the line $\cos 4\phi=-1$ is shifted to infinity;
the two satellites at $I=(b-a)/d$ 
have now angular coordinates $\cos 4\phi \approx -5$ and are therefore ghosts.
(Certainly they eventually may
become real at finite values of $\varepsilon$ and
$a$, far away from the codimension two bifurcation 
and therefore out of the scope of the present work.)
A tangent bifurcation is met at $\varepsilon=2a^2/(3c)$ provided that
$I=-a/(3c)>0$, since the Cartesian coordinates
(\ref{eq:canonpolarcoor}) are otherwise imaginary.
Sequences of a tangent bifurcation at positive $I$ and the
two variants of quadrupling bifurcations
are shown in Figure \ref{fig:m4cnt}.

\subsection{Period--$m$ bifurcation with $m\geq 5$
and tangent bifurcations of satellites} 

The codimension one bifurcations for $m\geq 5$ follow the 
island--chain pattern already encountered for $m=4$: There are two
satellites
that are ghosts on one side of the bifurcation and real on the other, 
one of them being stable and the other unstable. The stable and unstable
periodic points form a chain similar to the broken rational tori that
appear in almost integrable systems. Indeed, the $\phi$--dependent terms
in the normal forms are of the form of a small perturbation in that
situation.

\begin{figure}
\ifprep
\epsfxsize16cm
\epsfbox{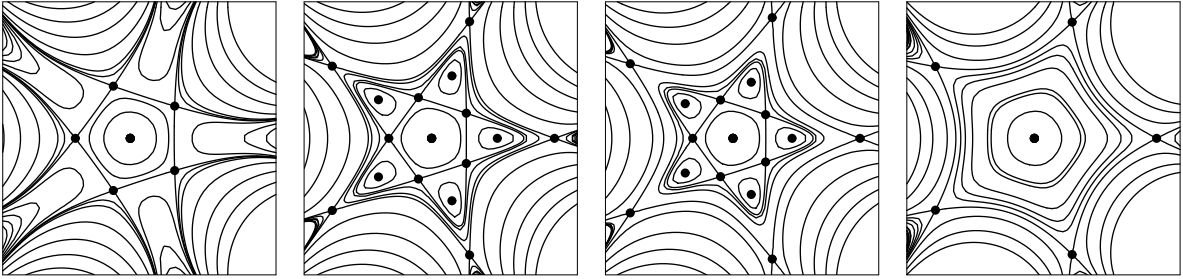}
\fi
\caption{
The bifurcation scenario close to the codimension two point 
with $m=5$, consisting of a tangent bifurcation of satellites
and a period--5 bifurcation
}
\label{fig:m5cnt}
\end{figure}

For $m\geq 5$ the usual Birkhoff normal forms describe
even the codimension two bifurcations: 
in addition to the orbits participating in the
period--$m$ bifurcation of codimension one they also account for the
satellites that are involved in the subsequent tangent bifurcations.
In the case $m=5$ one obtains 
three satellites at $\sin 5\phi=0$ that satisfy on the $p$--axis
\begin{equation}
\label{eq:m5fp}
\varepsilon +ap^2+\frac 5{4\sqrt 2} b p^3 =0\;.
\end{equation}
As for $m=1$ it is the discriminant
\begin{equation}
D=2\left(\frac 2 5\frac \varepsilon b\right)^2+\frac25
\frac \varepsilon b \left(\frac 8 {15}\frac a b\right)^3
\end{equation}
of the equation that governs
the number of real solutions. There is the period--5 bifurcation at
$\varepsilon=0$ and a tangent bifurcation at  $\varepsilon=-128a^3/(675
b^2)$. That sequence is depicted in Figure \ref{fig:m5cnt}.

\begin{figure}
\ifprep
\epsfxsize16cm
\ifprepgood
\epsfbox{m16.ps}
\else
\epsfbox{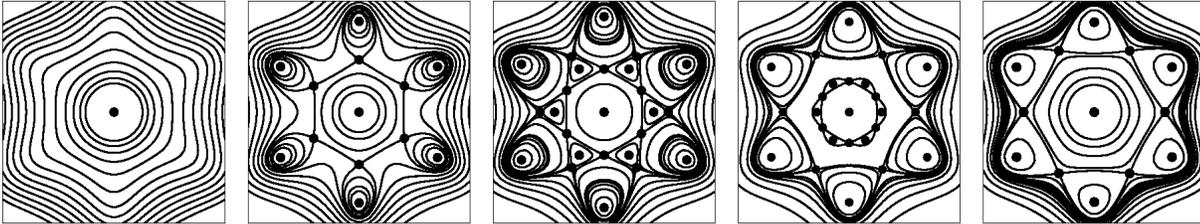}
\fi
\fi
\caption{Varying a parameter close to the codimension two point with 
$m=6$, one might observe
two pairs of satellites being born in a tangent bifurcation at positive $I$, as
displayed here; the
satellites closer to the center disappear in a subsequent period--6 bifurcation.
The island chain that is left over here
could be also steered to the center
by letting $a$, and then once more $\varepsilon$ change its sign.
}
\label{fig:m6cnt}
\end{figure}

In the case $m=6$ one finds
four satellites, two on each of the lines
$\cos 6\phi = \pm1\equiv \sigma$ at 
\begin{equation}
I^{(\sigma)}_\pm=-\frac 13 \frac{a}{b+\sigma c}\pm
\sqrt{\frac 19 \frac{a^2}{(b+\sigma c)^2}
-\frac 13\frac{\varepsilon}{b+\sigma c}}
\;.
\end{equation}
Tangent bifurcations
take place at independent parameter
combinations
$\varepsilon=a^2/[3(b+\sigma c)]$ (provided that the $I$--coordinate is
not negative).
A sequence with two tangent bifurcations at positive values of $I$ is
shown in Figure \ref{fig:m6cnt}.
All four satellites approach
the center in the codimension two bifurcation 
as $\varepsilon$ and $a$ are sent to zero.

\begin{figure}
\ifprep
\epsfxsize16cm
\ifprepgood
\epsfbox{m17.ps}
\else
\epsfbox{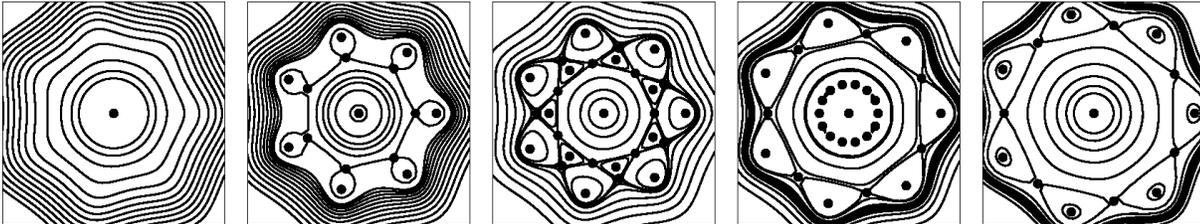}
\fi
\fi
\caption{
For $m\geq 7$ the resonant $\phi$--dependence of the normal form is weak,
and tangent
bifurcations of satellite pairs happen at almost identical values.
Subsequently the inner
island chain collapses onto the center and disappears. The remaining chain
might follow, as explained for $m=6$.
}
\label{fig:m7cnt}
\end{figure}

For $m\geq 7$ there are even more satellites in the Birkhoff normal form
than the four that participate in the codimension two scenario. 
In first order, the relevant satellites lie on the lines
$\cos m\phi=\pm1 \equiv \sigma$ at a radial distance
\begin{equation}
I^{(\sigma)}_\pm=
-\frac 13 \frac ab \pm \sqrt{\frac 19 \frac {a^2}{b^2} -\frac 13
\frac \varepsilon b}
\;,
\end{equation}
which is independent of $\sigma$.
The $\phi$--dependent
term induces a small correction of order $\sigma I^{m/2-2}$.
Before the tangent bifurcations, which are encountered
at almost identical values
$\varepsilon\approx a^2/(3b)$,
both satellite pairs have complex $I$. After the bifurcation
both the inner as well as the
outer orbits form island chains that are visible in phase space if
the $I$--coordinate is positive.
The corresponding sequence is shown in Figure \ref{fig:m7cnt}.

\section{Periodic orbits in semiclassical approximations}
\label{sec:deriv}

Individual contributions of periodic orbits to semiclassical expressions
of traces lose their validity close to bifurcations.
A collective
treatment of the involved orbits is then necessary.
Here we discuss general
aspects 
\cite{Gutzwiller:1971,Gutzwiller:1990,Tabor:1983,Junker:Leschke:1992,Ozorio:1988}
in which uniform approximations that provide the required regularization
are integrated in the next section.

Gutzwiller's trace formula relates the trace of the retarded Green's function 
\begin{equation}
G(E)=\frac{1}{E+i0^+-H}
\end{equation}
of an autonomous system with Hamiltonian $H$ 
to the properties of the periodic orbits in the classical 
system and thus provides a semiclassical expression of
the density of states
\begin{equation}
d(E)=-\frac1\pi \mbox{Im}\, \tr G(E)\;.
\end{equation}
A suitable starting point for the derivation of periodic--orbit contributions
is given by an integral
over Poincar{\'e} surfaces of section $\Omega$
(see e.\,g.\ \cite{Sieber:1996a}),
\begin{eqnarray}
\tr G(E)&=&\sum_{n=1}^{\infty}
\frac 1 {i\hbar} \int_{\Omega}\frac{\mbox{d}q_n\,\mbox{d}p_0}{2\pi\hbar}
\frac 1 n \frac {\partial \hat S^{(n)}}{\partial E}
\left|\frac{\partial^2\hat S^{(n)}}{\partial q_n\partial p_0}
\right|^{1/2}
\nonumber\\
&&\!\!\!\!
\times
\exp\left[\frac i\hbar [\hat S^{(n)}(q_n,p_0;E)-q_np_0]-i\frac\pi2
\nu\right]
\;.
\label{eq:intaut}
\end{eqnarray}
This involves the generating function $\hat S^{(n)}(q_n,p_0;E)$
of the $n$th iterate of the Poincar{\'e} map $(q_0,p_0)\to(q_n,p_n)$ with
\begin{equation}
\label{eq:saut}
\frac{\partial \hat S^{(n)}}{\partial q_n}=p_n\;,\qquad
\frac{\partial \hat S^{(n)}}{\partial p_0}=q_0\;,\qquad
\frac{\partial \hat S^{(n)}}{\partial E}=T\;,
\end{equation}
where $T$ denotes the time elapsed along the trajectory.
The integration domain $\Omega$ might be disjunct and is also used to
account for the different sheets of the generating function, i.\,e., the
multivaluedness of $\hat S^{(n)}(q_n,p_0;E)$. Finally, $\nu$ is the
Morse index.

A similar expression can also be obtained for periodically driven
systems that are described stroboscopically by a
time--evolution operator $F$.
For convenience we set the stroboscopic period to unity.
$F$ represents in general
the unitary operator of a quantum map with
a classical limit. The eigenstates of $F$ are stroboscopically
stationary, and the phases of the unimodular eigenvalues are called
quasienergies. The quasienergy spectrum is encoded in the traces $\tr
F^n$.
With the Van Vleck propagator
one obtains the expression
\begin{eqnarray}
\tr F^n&=&
\int_{\Omega}\frac{\mbox{d}q_n\,\mbox{d}p_0}{2\pi\hbar}
\left|\frac{\partial^2\hat S^{(n)}}{\partial q_n\partial p_0}
\right|^{1/2}
\nonumber\\
&&\times
\exp\left[\frac i\hbar [\hat S^{(n)}(q_n,p_0)-q_np_0]-i\frac\pi2
\nu\right]
\label{eq:intmap}
\end{eqnarray}
where $\hat S^{(n)}(q_n,p_0)$ is now the generating function of the $n$--step
map with
\begin{equation}
\frac{\partial \hat S^{(n)}}{\partial q_n}=p_n\;,\qquad
\frac{\partial \hat S^{(n)}}{\partial p_0}=q_0\;,
\end{equation}
and the integration domain $\Omega$ lies in phase space
[again, accounting also for the multivaluedness of $\hat S^{(n)}(q_n,p_0)$].

We consider contributions
to $\tr F^n$ or $i\hbar \tr G(E)$, the factor  being
introduced to facilitate a parallel investigation of both cases.
Let us examine the contribution of an
arbitrarily chosen region $\Omega'$,
\begin{equation}
\label{eq:pqint}
C_{\Omega'}=\frac 1 {2\pi\hbar} 
\int_{\Omega'}\mbox{d}q'\;\mbox{d}p\;\Psi(q',p)\exp\left[\frac i \hbar 
\Phi(q',p)-i\frac \pi 2 \nu\right]
\;,
\end{equation}
with the notations $q'\equiv q_n$, $p\equiv p_0$.
For the contributions to $i\hbar \tr G(E)$
the phase function $\Phi$ and the amplitude function $\Psi$ are
\begin{eqnarray}
\Phi(q',p)&=& \hat S^{(n)}(q',p;E)-q'p\\
\Psi(q',p)&=& \frac {1} n \frac{\partial \hat
S^{(n)}}{\partial E}\left|\frac{\partial^2 \hat S^{(n)}}{\partial
q'\partial p}\right|^{1/2}
\;.
\end{eqnarray}
For contributions to $\tr F^n$ they read
\begin{eqnarray}
\Phi(q',p)&=& \hat S^{(n)}(q',p)-q'p\\
\Psi(q',p)&=&
\left|\frac{\partial^2\hat S^{(n)}}{\partial
q'\partial p}\right|^{1/2}
\;.
\end{eqnarray}
The main contributions to the integral (\ref{eq:pqint})
arise from regions $\Omega'$ around the stationary points. These are
given by $\frac{\partial \Phi}{\partial q'}=0$ and $\frac{\partial
\Phi}{\partial p}=0$ which corresponds to
\begin{equation}
\label{eq:statphase}
\frac{\partial\hat S^{(n)}}{\partial p_0}=q_n\;,
\frac{\partial\hat S^{(n)}}{\partial q_n}=p_0\;.
\end{equation}
The solutions are the periodic points of a given energy
in the autonomous case or given stroboscopic period in the driven case.
These regions come naturally into focus when one uses the
steepest--descent method to find the leading--order term of an
asymptotic expansion of the integral in $\hbar$.
To achieve this goal the integration variables are complexified and the
initial contour is deformed 
such that the maxima lie at the solutions of (\ref{eq:statphase}).
From the maxima one follows paths of steepest descent of the
integrand.
The new contour has to originate from the original one by a
continuous deformation without crossing singularities. 
In order to construct a contour that connects the original
integration boundaries one automatically visits also some complex `ghost'
solutions of (\ref{eq:statphase}).
Only the ghosts that lie on the deformed contour are relevant.
We come back to this point in the discussion of global bifurcations
in section \ref{sec:stokes}.

The size of the region $\Omega'$ of almost stationary phase
depends on $\hbar$ and
is determined by the condition that the typical variation of the phase
$\Phi(q',p)/\hbar$
over that region is of order $2\pi$.
The region can contain several stationary points, i.\,e., phase space
points of a number of periodic orbits, and each orbit may contribute
with several points.

The region shrinks if $\hbar$ is sent to zero
while all parameters are fixed at general values (codimension zero),
and finally splits into regions containing only a single periodic point.
Each of these gives rise to a usual stationary--phase (sp) contribution
\begin{equation}
\label{eq:statphaseint}
C_{\Omega'}^{(\rm sp)}=
\frac{T_0}{n_0}
\frac {\exp\left[\frac i\hbar S^{(n_0r)}-i\frac \pi2\mu\right]}{|2-\tr M^{(n_0r)}|^{1/2}}
\;.
\end{equation}
There are $n_0$ such contributions from each orbit.
Moreover, repetitions of an orbit are regarded as independent here.
Four canonical invariant
characteristic quantities of periodic points
enter this semiclassical expression, the primitive period
$T_0$, the action $S^{(n)}$,
the stability factor $\tr M^{(n)}$, and the Maslov index $\mu$.
They also determine the uniform approximations to be derived in the next
section.

In the periodically driven case, $T_0=n_0$ is the primitive stroboscopic
period that was introduced in section \ref{sec:nform}.
(Recall that we set the stroboscopic period to unity.)
The action $S$
is given by the value of
Hamilton's
principal function, $S=\int (p\,\mbox{d}q-H\,\mbox{d}t)$, or, equivalently, by the value
of the phase
\begin{equation}
\label{eq:action}
S=\hat S^{(n)}(q,p)-q p
\end{equation}
at the periodic point.
The linearized $n$--step map $M^{(n)}$ was likewise introduced in
section \ref{sec:nform}. It
is connected to the second derivatives of $\hat S^{(n)}$
and involved in the expression through its trace
\begin{equation}
\label{eq:trM}
\tr M^{(n)}=
\frac{
1 +
\left(\frac{\partial^2 \hat{S}^{(n)}}{\partial q' \, \partial p}\right)^2 -
\frac{\partial^2 \hat{S}^{(n)}}{\partial q'^2}
\frac{\partial^2 \hat{S}^{(n)}}{\partial p^2}
}{
\frac{\partial^2 \hat{S}^{(n)}}{\partial q' \, \partial p}
}
\;.
\end{equation}
Finally, there is the Maslov index $\mu=\nu -\frac 12 \mbox{sign}\, G$
where $\nu$ is the Morse index and
$\mbox{sign}\, G$ denotes the difference in the number of positive and
negative eigenvalues of the matrix
\begin{equation}
\label{eq:gmat}
G=\left(\begin{array}{cc}
\frac{\partial^2 \Phi}{\partial q'^2} &
\frac{\partial^2 \Phi}{\partial q'\partial p} \\[.2cm]
\frac{\partial^2 \Phi}{\partial q'\partial p} &
\frac{\partial^2 \Phi}{\partial p^2}
\end{array}
\right)
\;,
\end{equation}
involving second
derivatives of $\Phi=\hat S^{(n)}-q'p$.

The four quantities are almost of the same form in the
autonomous case.
The 
primitive period $T_0$ is the smallest period $T$ [see eq.\
(\ref{eq:saut})]
after which one comes back to the initial point when starting somewhere
on the trajectory of the periodic orbit.
The orbit shows up as a fixed point in all
$n$--step maps on the surface of section where, as before,
$n=rn_0$, and $r=T/T_0$ is an integer
counting repetitions.
The quantity $S$ is again given by (\ref{eq:action}) but corresponds now
to the reduced  action
$S=\oint p\,\mbox{d}q$.
The linearized map on the surface of section and the
stability factor are connected to the generating function in the same
way as before,
and the Maslov index
is again connected to the Morse index and the matrix (\ref{eq:gmat}) by
$\mu=\nu -\frac 12 \mbox{sign}\, G$.

\section{Uniform approximations}
\label{sec:unif}
\subsection{Breakdown of stationary phase near bifurcations}
From the bifurcation condition (\ref{eq:bifcond}) it follows
that the individual contribution (\ref{eq:statphaseint})
of an orbit blows up close to a bifurcation
and even diverges right at $\tr
M=2$.
The assumption under which 
a stationary--phase approximation is reasonable is then no longer
fulfilled:
It does not suffice to expand the generating function up to second order
around the trajectory,
\begin{equation}
\label{eq:phistart}
\Phi(q',p)=S_0-\frac \omega 2q'^2-\frac \sigma 2 p^2 +{\cal
O}(3)
\;,
\end{equation}
since the coalescing orbits cannot be separated even in the limit
$\hbar\to 0$.
In the two--parameter family of Hamiltonians (\ref{eq:hamfam})
the stationary--phase
approximation is in danger 
close to bifurcations of codimension one
and two.  In practical applications with
small but finite values of $\hbar$, one typically even visits regions in
which bifurcations of higher codimension are felt.
Along the same line, bifurcations of codimension two are
even felt when only a single
parameter (frequently, none at all) is varied;
this observation is
indeed our principal incentive.

In this section we derive normal forms for 
the phase function $\Phi$ and the amplitude function $\Psi$
in (\ref{eq:pqint})
that supersede the quadratic form (\ref{eq:phistart}) and
yield regular expressions
as a substitute for the stationary--phase result (\ref{eq:statphaseint})
in regions affected by bifurcations of codimension two.

\subsection{Normal forms for phase and amplitude}

To each normal form of the Hamiltonian there is a corresponding expression
for the generating function which carries over to the phase function $\Phi$.
This function has as many stationary points as the 
Hamiltonian. The simplest functional form that can be achieved 
is identical to the normal form of the Hamiltonian, but with altered
coefficients. (Observe, however,
that although this form of
$\Phi(q',p)$, when expressed by 
canonical polar coordinates, obeys again rotational symmetries $C_m$, 
this is no longer the case for the map generated by it.)
To illustrate this identification
we note that the normal forms effectively
describe the integrable dynamics
of an autonomous system with one degree of freedom. 
In action--angle variables $J$, $\psi$ the evolution over a time
interval of duration one is
generated by
\begin{equation}
\label{eq:acham}
\hat S(J,\psi')=S_0+(\psi'+2\pi n) J-H(J)\;,
\end{equation}
such that $J=J'$ and $\psi=\psi'-\omega(J)\bmod 2\pi$ with the torus
frequency $\omega(J)=\mbox{d}H/\mbox{d}J$.

The action variable $J$ is quantized since
the angle $\psi$ is only given modulo $2\pi$; this gives rise to
branches of the generating function that are here enumerated by
$n$.
We circumvent this obstacle by considering the map in Cartesian coordinates,
generated by $\hat S(q',p)$, and appealing to
the canonical invariance of the leading--order term in the
$\hbar$--expansion that we are looking for.
[This does not affect the option to shift finally to canonical polar coordinates
in the integral expression (\ref{eq:pqint}).]
The transformation to these
variables yields new coefficients: for instance,
while the second derivatives
of $H$ around a stable orbit involve the bare stability angle $\omega$,
we are led for the generating function
to the relation (\ref{eq:trM}) with $\tr M=2\cos\omega$.
The coefficients in $\Phi$ are therefore related,
but not identical to those in the normal
forms of the Hamiltonian, although we will not reflect this in a change
of notation; however,
minus signs will be introduced following a convention that is
motivated by eq.\ (\ref{eq:acham}).

In section \ref{sec:nform} we observed that the normal forms 
describe in some cases additional orbits that do not
participate in the bifurcations of codimension two. In addition
we will see that there are
in general not enough coefficients to yield independent actions and
semiclassical amplitudes in the stationary--phase approximation. To overcome
these restrictions one has to consider the influence of higher--order terms in
the normal forms on the classical properties of the orbits. These
terms can be eliminated in the region of almost stationary
phase by non--canonical transformations. Such
transformations can also be used to get rid of the additional non--bifurcating
orbits described by the original normal form.
The Jacobian of the
transformation enters the amplitude function $\Psi$, which can be
simplified further by partial integrations.
The procedure is sketched in section \ref{sec:unifderiv}.
It results in the following
normal forms for $\Phi$ and $\Psi$,
\begin{equation}
\begin{array}{|c|l|}
\hline
m & \Phi^{(m)}(q',p)-S_0  \\
\hline
1 & - \varepsilon q'^2 - a q'^3 - b q'^4 - \frac \sigma 2 p^2\\
2 & -\varepsilon q'^2 - a q'^4 - b q'^6 - \frac \sigma 2 p^2\\
3 & -\varepsilon I - a I^{3/2}\cos 3\phi - b I^2 \\
4 & -\varepsilon I - a I^2(1+\cos 4\phi) - b I^2(1-\cos 4\phi) 
\\ & \qquad -c I^3 (1+\cos 4\phi)\\
5 &   - \varepsilon I - a I^2 - b I^{5/2}\cos 5\phi - \varepsilon
c I^{3/2}\cos 5\phi\\
6 &   - \varepsilon I - a I^2 - b I^3 - c I^3\cos 6\phi
-\varepsilon d I^2 \cos
6\phi\\
\geq 7,\,\mbox{odd} &  - \varepsilon I - a I^2 - b I^3 
-   I^{3/2}(c I+\varepsilon d) \cos m\phi 
\\
\geq 8,\,\mbox{even} &  - \varepsilon I - a I^2 - b I^3 
-  I^2(cI+\varepsilon d)\cos m\phi
\\
\hline
\end{array}
\;,
\end{equation}
\begin{equation}
\begin{array}{|c|l|}
\hline
m & \Psi^{(m)}(q',p) \\
\hline
1 &
 1+ \alpha q'+\beta q'^2 \\
2 &
 1+ \alpha q'^2+\beta q'^4 \\
3 &
 1+ \alpha I +\beta I^2 \\
4 &
 1+ \alpha I +\beta I^2 + \gamma I^2\cos 4 \phi + \delta I^3 \\
5 &
 1+(\alpha+\beta I) I^{1/2} \cos 5\phi + \gamma I   +\delta I^2 \\
6 &
 1+(\alpha +\beta I)I\cos 6\phi+ \gamma I + \delta I^2 + \xi I^3 \\
\geq 7,\,\mbox{odd}
 & 1+(\alpha+\beta I) I^{1/2} \cos m\phi + \gamma I   +\delta I^2+\xi I^3 \\
\geq 8,\,\mbox{even}
 & 1+(\alpha +\beta I)I\cos m\phi+ \gamma I + \delta I^2 + \xi I^3 \\
\hline
\end{array}
\;.
\end{equation}
In $\Phi^{(1,2)}$ we demand  $|\sigma|=1$.
The normal forms for $m\geq 3$ are expressed in
canonical polar coordinates $I$, $\phi$ defined in
equation (\ref{eq:canonpolarcoor}).
Note that 
terms show up in the expressions
for $m\geq5$ that cannot be expressed `perturbatively' as $q'^lp^k$ in 
Cartesian coordinates (\ref{eq:canonpolarcoor}).
The substitutions $m\phi= 2\psi$ for
$m$ even and $m\phi=\psi$ for $m$ odd
provide a potentially useful regularization.

Collective contributions of 
bifurcating periodic orbits 
are obtained by inserting the normal forms for $\Phi$
and $\Psi$ into (\ref{eq:pqint}).

Numerically useful expressions
in terms of Taylor series can be found for some of the integrals in
\cite{Main:Wunner:1997} ($m=1,2$) and \cite{Schomerus:1997a} ($m=3$).
The integrals for $m=1,2$ can also be easily
evaluated by the method of
steepest descent (cf.\ section \ref{sec:stokes})
since they are essentially one--dimensional. For $m\geq 3$, however,
a two--dimensional
steepest--descent manifold might be quite difficult to construct.
It is perhaps more convenient to deform only the $I$ coordinate into the
complex, yielding a simple
steepest--descent contour for each fixed, real $\phi$, 
and then to perform the $\phi$--integral of finite range.

\subsection{Determination of coefficients}

The coefficients in the normal forms have to be expressed by the
classical quantities $S$, $\mbox{tr}\,M$
of the orbits in order to obtain
a contribution that is invariant under canonical transformations
\cite{Sieber:1996a,Schomerus:Sieber:1997a,Sieber:Schomerus:1997a,%
Tomsovic:Grinberg:1995,Ullmo:Grinberg:1996}.
This is achieved by examination of the stationary--phase limit 
$\hbar\to 0$ of (\ref{eq:pqint}) while all other coefficients are fixed.
With definition (\ref{eq:gmat}),
each stationary point gives rise to a contribution
\begin{equation}
C^{(\rm sp)}=\frac{\Psi}{\sqrt{\left|\mbox{det}\,G\right|}}\exp\left[\frac i\hbar
\Phi-i\frac \pi 2
\left(\nu-\frac 12 \mbox{sign}\,G\right)\right]
\;,
\end{equation}
which determines the parameters by comparison to (\ref{eq:statphaseint}).
The collective contributions are then not only applicable in the
immediate
neighbourhood of the bifurcation, but also far away, where they split
into a sum of isolated contributions of Gutzwiller type
(\ref{eq:statphaseint}) with correct amplitudes and phases.
Consequently they constitute uniform approximations.
(Transitional approximations of the
type mentioned in the introduction are obtained if one uses $\Psi=1$
instead.)

In detail,
the properties of the central orbit determine $\varepsilon$ and $S_0$,
since the stationary point in the origin gives
$C^{(\rm sp)}=\exp\left[iS_0/\hbar-i\frac\pi 2\left(\nu+\frac 12 (\sigma+\mbox{sign}\,
\varepsilon)\right)\right]/\sqrt{|2\varepsilon|}$ for $m=1,2$ (recall
that $|\sigma|=1$) and
$C^{(\rm sp)}=\exp\left[iS_0/\hbar-i\frac\pi 2(\nu+\mbox{sign}\,
\varepsilon)\right]/|\varepsilon|$ for $m\geq 3$. 
The remaining coefficients of the phase function are uniquely determined
by the actions of the satellites.  It turns out to
be helpful to use an ansatz
where the coefficients are expressed by scaled positions on a radial line connecting the satellites with the central orbit.
For even $m$ with two real satellites
on such a line, for instance, we
put them on scaled positions at $x_1=\pm1$, $x_2=\pm y$, corresponding to
\begin{equation}
\frac{\mbox{d}\Phi}{\mbox{d}x}=Ax(x^2-1)(x^2-y^2)
\;,
\end{equation}
integrate to obtain $\Phi(x)$
and determine $y$ from the (scale invariant) ratio 
\begin{equation}
\frac{S_1-S_0}{S_2-S_0}=\frac{1-3 y^2}{y^4(y^2-3)}
\;.
\end{equation}
Without restriction we can demand $0\leq y\leq 1$; then there is exactly
one solution.
The factor $A$ follows from the absolute value of $S_1-S_0$, and the scale
of $x$ is fixed by knowledge of $\varepsilon$.
In the case of
complex satellites they are placed at $x=\pm 1\pm i y$, i.\,e.,
$\Phi=Ax^2[x^4+3x^2(y^2-1)+3(y^2+1)^2]$,
and $y$ is obtained from
\begin{equation}
\frac{\mbox{Re}\,S_1-S_0}{\mbox{Im}\,S_1}=\frac{1+9 y^2-9 y^4-y^6}{16 y^3}
\;.
\end{equation}
There is a solution with $|y|<1$ and another one with $|y|>1$.
The right choice takes into consideration
whether the ghost with $\mbox{Im}\,S>0$
lies on the steepest--descent contour connecting the integration
boundaries or not; see section \ref{sec:stokes}.

The approach presented here to obtain the coefficients of $\Phi$ works
also for the other normal forms.
Moreover, the stationary--phase result is a linear
combination of 
the coefficients of the amplitude function $\Psi$, which are therefore
easily obtained by comparison with the semiclassical amplitudes
in (\ref{eq:statphaseint}).
For $m\geq 5$ there is a symmetry--related pair of
`spurious' ghost satellites (analogous to
those already discussed for $m=4$), which are negligible not only since
they do not lie on a steepest--descent contour (see again
section \ref{sec:stokes}),
but also because a `magic' cancellation in course of the derivation (see
below) entails $\Psi=0$ at their positions.
Fortunately,
exactly one extra term in $\Psi$ shows up in these
cases which can be used to
achieve this suppression.
It seems reasonable that one uses this approach also in
the case $m=4$, where an extra coefficient is also at one's
disposal.

\subsection{Derivation of normal forms}
\label{sec:unifderiv}

We perform now the reduction of the original normal forms for $\Phi$
(identical in appearance to those of $H$) to the forms listed above
and obtain in
parallel the expressions of $\Psi$.
The remainder consists of higher orders in $I$ as well as 
$a$, $\varepsilon$.
Once more we invoke 
Vieta's relations and regard the coefficients as certain orders of the typical
distance of 
the satellites to the
central orbit. The coefficient  $\varepsilon$
is usually the product of such
distances, while $a$ is a sum.
We use 
that the original normal forms entail
$\Psi=\mbox{const}$ plus corrections of the form $q'^lp^k$.
In most cases the precise
form of
coefficients in the transformations
is of no particular interest
and therefore not given;
the expressions are easily obtained, for instance, with the 
assistance of symbolic mathematical programs.
Only the feasibility of reduction counts.
Finally all coefficients in $\Psi$ and $\Phi$ are not to be determined
from explicit expansions of Hamiltonians or generating functions but
rather from the actions and stability properties of the orbits as
explained above; otherwise no canonically invariant result would be
obtained.

For $m=1,2,3$ all the periodic points described by
the original normal forms are involved in the bifurcation.
For $m=1$ there is at least one real orbit which we place
into the origin
by a shift of $q'$. This results in the normal form $\Phi^{(1)}$. 
The three coefficients $\varepsilon$, $a$, and $b$ as well as the value
$S_0=\Phi^{(1)}(0,0)$ are fixed by the three actions of the orbits
and one of the stability factors. From the normal form follows
$\Psi=1$, the other stability factors are
therefore not yet independent. At a little
distance to the bifurcation,
however, higher--order terms in the Hamiltonian or the
phase function
act as a perturbation, and
the implied relations between the classical quantities of the orbits
are no longer valid. For $m=1$ we consider
terms of type $cq'^5+dq'^6$. Terms involving $p$ effectively
do not alter the final expression and can be discarded by appealing to the
splitting lemma of catastrophe theory.
In the region of almost stationary phase the higher--order
terms act as a perturbation and
can be eliminated
by substituting $q'=Q+AQ^2+BQ^3$ with suitably chosen
coefficients. The Jacobian of the transformation 
involves $\mbox{d}q'=\mbox{d}Q(1+2AQ+3BQ^2)$ and gives 
the normal form $\Psi^{(1)}(q',p)$ announced above. The two
additional coefficients are determined by the remaining
stability factors $\mbox{tr}\,M$ of the satellites.
Corrections to $\Phi$ of even higher order would carry over
to higher--order terms in $\Psi$. They involve additional
coefficients and on first sight allow for ambiguities, but
can be eliminated by successive partial integrations. The term of
highest order is written as
\begin{equation}
Q^{l}\sim Q^{l-3}\frac{\mbox{d}\Phi^{(1)}}{\mbox{d}Q}
+\mbox{terms of order $l-1$,
$l-2$}
\;.
\end{equation}
The partial integration of the first term gives an order $\hbar
Q^{l-4}$ and a boundary contribution that vanishes for $l\geq 4$.
In the course of this procedure
the constant $1$ and the coefficients $\alpha$ and $\beta$ in $\Psi^{(1)}$
acquire 
next--to--leading order corrections in $\hbar$ that can be discarded and
reflect canonically
non--invariant properties of the orbits.

For $m=2$ the higher--order terms have to obey the reflection symmetry
$q'\to -q'$ and are on the $q'$--axis of the form $d q'^8+ e q'^{10}$. 
They are eliminated by $q^2=Q^2+AQ^4+BQ^6$. The Jacobian involves only
even orders of $Q$. Terms of order $l\geq 6$ can be eliminated in $\Psi$
by writing
\begin{equation}
Q^{l}\sim Q^{l-5}\frac{\mbox{d}\Phi^{(2)}}{\mbox{d}Q}
+\mbox{terms of order $l-2$, $l-4$}
\end{equation}
and performing a 
partial integration of the first term that yields a term $\sim \hbar
Q^{l-6}$.
This reduces the amplitude function to its normal form $\Psi^{(2)}$.

In the case $m=3$ we have to consider the influence of
the terms $c I^{5/2}\cos 3\phi+d I^3$.
We can safely use $\varepsilon={\cal O}(I^*)$ and $a={\cal O}({I^*}^{1/2})$ as
upper bounds in orders of the typical distance $I^*$ of the satellites to the
center, which is in turn connected with the size of the region of almost
stationary phase.  Note that this does not impose a restriction on the
relative order of these parameters as long as they are small enough.
Having this in mind, it is easy to see that
the elimination of the extra term can be  achieved
by a transformation
$I=J+A J^{3/2}\cos 3\psi$ and $\phi=\psi+B J^{1/2}\sin 3\psi$. The
corresponding coefficients $A$ and $B$ give rise to a
cancellation of the order $J^{1/2}$ in the Jacobian. This order also does
not show up in the original $\Psi$, since it can expressed by terms
of the form $q'^lp^k$ and obeys a three-fold symmetry.
Hence, the order $I^{1/2}$ is absent in $\Psi^{(3)}$. Higher orders are
again eliminated by partial integrations, carried out now with respect to
$I$. The integration over $\phi$ suppresses 
terms that are odd in $\phi$, such as $\sin 3\phi$.
Note that one
could work alternatively with the form $\tilde\Psi^{(3)}=1+\alpha I+\beta
I^{3/2}\cos 3\phi$; the equivalence to $\Psi^{(3)}$
is again worked out by a partial integration. 

Only three of the six satellites described by  $\tilde H^{(4)}$
(\ref{eq:h4tilde})
are involved in the
codimension two bifurcation with $m=4$. The normal form  
$\Phi^{(4)}$ given above has been given a deeper foundation
in the discussion of
the corresponding Hamiltonian $H^{(4)}$.
The outer satellite
on the lines $\cos 4\psi=-1$ is shifted to infinity and is infinitely
unstable, $|\tr M|=\infty$. The two satellites
at radial distance $I=(b-a)/c$ have $\cos 4\psi\approx -5$.
They are consequently
ghosts with real actions and do not contribute in the stationary--phase limit
(for a deeper foundation see section \ref{sec:stokes});
moreover, they are quite far away from the bifurcating orbits. We
assume for that reason that their influence is negligible.
This leaves us with a normal form that is again completely determined by the
actions and one stability factor. We assume that the three non--bifurcating
satellites remain negligible even under the influence of higher--order
terms.
The corresponding amplitude normal form $\Psi^{(4)}$ is irreducible
under further partial integrations, but has one more
coefficient than needed to account for independent stability factors of
the satellites. In analogy to the situation to be discussed for $m\geq
5$ we can use this coefficient to yield $\Psi=0$ at the
position of the unwanted ghosts in favor of an additional suppression.

For $m=5$ there are three bifurcating satellites but not enough coefficients
in the original normal form to account for independent actions.
Observe
that a scaling transformation $I=AJ$ does not affect the values of the
phase function at the stationary points.
Only the two 
combinations $\varepsilon^2/a$ and $\varepsilon^{5/2}/b$
enter these values which have to match the three actions
of the satellites.
(The coefficients are determined uniquely if one
takes the stability factors into consideration.)
Independent actions are admissible after allowing for higher--order
terms which are once more removed by a transformation in order to yield no
spurious additional stationary points. The next--order term,
conveniently expressed as $+5bc/2 I^3$, can be
eliminated by a transformation 
\begin{eqnarray}
I&=&J+c J^{3/2}\cos5 \psi\\
\phi&=&\psi-\frac c2 J^{1/2}\sin 5\psi
\end{eqnarray}
which is similar to the one
for $m=3$,
but now the order $J^{1/2}$ survives in the Jacobian,
\begin{equation}
\label{eq:psi5}
\Psi=1-cJ^{1/2} \cos 5\psi+{\cal O}(J)
\;.
\end{equation}
The transformation gives rise to the term $-\varepsilon c J^{3/2}\cos 5\psi$
in $\Phi$ and provides us with the additional coefficient $c$.
[For illustrative purposes we wish to mentioned here that the
coefficient $\varepsilon c$ in $\Phi$ can be treated as
${\cal O}({J^*}^{3/2})$,
where $J^*$ gives the order
of the distance of the satellites to the central orbit.
This order is related
to the coefficients $\varepsilon$ and $a$ by application of Vieta's
relations to the stationary--point equation $\partial \Phi/\partial I=0$.]
The
next--order corrections to $\Phi$ give even one coefficient more in $\Psi$
than necessary for independent stability factors of the bifurcating 
orbits. There is, however, an extra pair of satellites at
$I=I^{(0)}\equiv-\varepsilon c/b$ that even approaches the center as $\varepsilon\to
0$. The angular coordinate, however, obtains a large imaginary part since
it obeys $ (I^{(0)})^{1/2}
c \cos 5\phi=1$, where a
term of order $a$ has been dropped.
Indeed this yields in leading order (\ref{eq:psi5}) a vanishing $\Psi$ and
encourages us to use
the extra
coefficient to accomplish suppression of the unwanted ghosts.

A similar situation is encountered 
for $m=6$: Only three independent actions of
four bifurcating satellites can
be modeled with the original normal form, but higher--order terms 
give rise to corrections that lead to the given normal forms of phase
and amplitude.
An extra coefficient is again present to suppress the ghost
pair at $I=-\varepsilon d/c$, and the Jacobian of the transformation
turns out to be once more in favor of such a strategy.

Enough coefficients for the actions are 
present in the original normal form of $\Phi$
for $m\geq 7$, but the expression yields more stationary points
than desired. The $\phi$--dependent terms 
of highest order in $I$ can be eliminated in favor of terms of lower
order by successive substitutions $I=J+AJ^l\cos m\phi$.  This procedure
can be carried out in parallel to substitutions $I=J+BJ^l$ that aim at
the elimination of $\phi$--independent terms. The coefficients of the
remaining terms reflect the values at the stationary
point up to the order that yields them as independent from each other
and allows also for independent stability factors through the expression
for $\Psi$. In the derivation one encounters again a Jacobian that
vanishes at the location of the unwanted ghost pair at $I=-\varepsilon
d/c$. We should note that the
coefficient $c$ is here of
order $a$, or, equivalently, $\varepsilon^{1/2}$.

\section{Stokes transitions}
\label{sec:stokes}
\subsection{Preliminary remarks}

We mentioned in section \ref{sec:deriv}
that a steepest--descent contour
has also maxima in complexified phase space which correspond to ghost
solutions of the stationary--point equations.
A helpful rule in that respect is that
only ghosts with $\mbox{Im}\, \Phi >0$ can lie on the steepest--descent
contour.
Moreover,
satellites that disappear in period--$m$ bifurcations of codimension
one with even $m$ are afterwards `self--conjugated' ghosts,
that is, map onto
themselves under complex conjugation, and have therefore real
classical quantities (in canonical polar coordinates they have real
$\phi$ and $I<0$); accordingly, they do not contribute.
The ghosts immediately beyond a
period--$m$ bifurcation of codimension one with odd $m\geq 5$ have
almost real action, a small imaginary part only being introduced
from higher orders, and also do not contribute. 
For the remaining ghosts, however, it cannot be avoided to construct
the contour in order to find
out whether they are relevant or not
(although the majority of relevant ghosts will be close to reality, i.\,e.,
about to bifurcate).

A steepest--descent contour consists of different sheets.
The phase $\mbox{Re}\, \Phi$ in the exponent of the
integrand is constant on each sheet
and thus given by its value at the periodic point,
that is, the real part of its action.
For general combinations of the control parameters (codimension zero)
there will be only one orbit lying on each sheet, though it is possible
that it does so with
more than one of the points along its trajectory.
Imagine now that for one combination of the parameters
a ghost
lies on the steepest--descent contour while for another one it
does not.
The ghost is denoted by $+$ in the following and sometimes 
called `subdominant' orbit. Its complex conjugate partner is denoted by
$-$.
Somewhere on a path connecting both parameter combinations
the contour changes its form
qualitatively in a so--called Stokes transition:
The sheet
of the ghost
merges with the sheet of another
orbit, which is called the `dominant orbit' and denoted by $0$. 
On both sides of the
transition the sheets of the two orbits will connect different zeroes
of the integrand. In the transition the contour changes in such a way
that the sheet of the subdominant orbit is no longer needed to connect
the original integration boundaries.

A necessary condition that the sheets of two orbits
merge is that the real part of
their actions are identical,
\begin{equation}
\label{eq:stokescond}
\mbox{Re}\,S_0=
\mbox{Re}\,S_+
\;.
\end{equation}
In general, this condition is not sufficient since both sheets
could be separated by others.
(Investigating energies of real orbits, the analogous condition
$H_0=H_+$ is necessary to find a heteroclinic orbit; again, additional
insight is needed to decide
whether an equipotential contour joining both orbits indeed exists.
These two global bifurcation types are therefore intimately related and 
constitute, in the language of
catastrophe theory, instances of saddle connections.)

\subsection{Stokes transitions in the diffraction integrals}

In the neighbourhood of codimension two bifurcations 
one encounters
tangent bifurcations in which satellites become ghosts, as was
demonstrated in
section \ref{sec:discn}.
Subsequently the ghosts may undergo Stokes transitions with the central orbit.
We analyze the transitions
by investigating the expressions for the actions
of the central orbit and the ghosts
and using the condition (\ref{eq:stokescond}).
We will demonstrate 
that no connection can exist between the sheet of the central orbit
and ghosts that have (almost) real $I<0$ and real actions.
Moreover
it is not difficult to realize that 
the condition (\ref{eq:stokescond}) is on the other hand
sufficient for ghosts beyond tangent bifurcations at positive $I$.
The reasoning is
facilitated by the observation that
the problem can be reduced  in all cases
to one dimension by considering the $I$--lines
(or, to be precise, the complex $I$--planes)
$\phi \bmod \pi =\mbox{const}$ that connect
the central orbit and the ghost satellites radially.

\begin{figure}
\ifprep
\epsfxsize5cm
\epsfbox{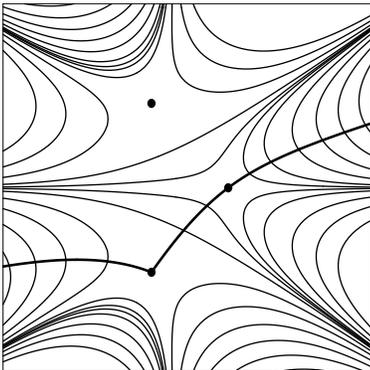}
\fi
\caption{
Path of steepest descent $\mbox{Re}\,\Phi^{(1)}(q,0)=S_0$ (thick line).
The
parameters are chosen to fulfill the condition for a Stokes transition. The transition
indeed takes place since the contour connects the subdominant ghost with the dominant
central orbit. The thin lines are the equipotential lines of $|\exp[i\Phi]|$
(or $\mbox{Im}\,\Phi$).
}
\label{fig:pver11cont}
\end{figure}

For $m=1$ and $m=3$ the situation is simple since
there are no other orbits than the two satellites (which are the ghosts
in question)
and the central orbit.
The diffraction integral for $m=1$ involves Pearcey's integral (and its
derivatives) for which the
Stokes transitions have been studied by
Wright \cite{Wright:1980}.
In the construction of $\Phi^{(1)}$ the real solution
has been placed in the
origin, $q'_0=p_0=0$. The satellites
have
coordinates $p_\pm=0$,
\begin{equation}
q'_\pm=-\frac 3 8 \frac a b\pm \sqrt{\frac 9 {64}\frac{a^2}{b^2}-\frac 12
\frac \varepsilon b}
\;.
\end{equation}
(A tangent bifurcation 
is now encountered at $\varepsilon=9a^2/32b$.
Two orbits coalesce
also at $\varepsilon=0$, but both remain real there due to the
construction.)
The orbit at the origin has the action $S_0$, while
\begin{equation}
S_\pm=S_0-{q'}_\pm^2\left(\frac\varepsilon 2+\frac a 4 q'_\pm\right)
\end{equation}
for the satellites.
The Stokes transition takes place at
\begin{equation}
\label{eq:m1stokes}
\varepsilon=\frac 3 {16}(3+\sqrt 3)\frac {a^2}b
\;.
\end{equation}
Figure \ref{fig:pver11cont}
displays the integration contour in the complex $q'$--plane for
$\varepsilon=3(3+\sqrt{3})/16$, $a=b=1$
together with the
equipotential lines of $|\exp[i\Phi]|$ (or, equivalently, of $\mbox{Im}\,\Phi$).
The plot demonstrates the well--known
existence of the connection and is characteristic even for arbitrary
sets of parameters that fulfill (\ref{eq:m1stokes}) since
the shape of the contour is fully determined by the combination
$\varepsilon b/a^2$:
The contour expands linearly with a scaling of $q$, such that we can
achieve, for instance, $a=b$, and does not change
if $\Phi^{(1)}$ is multiplied by a real constant, which allows to set $b=1$. 

\begin{figure}
\ifprep
\epsfxsize16cm
\ifprepgood
\epsfbox{m2cnta.ps}
\else
\epsfbox{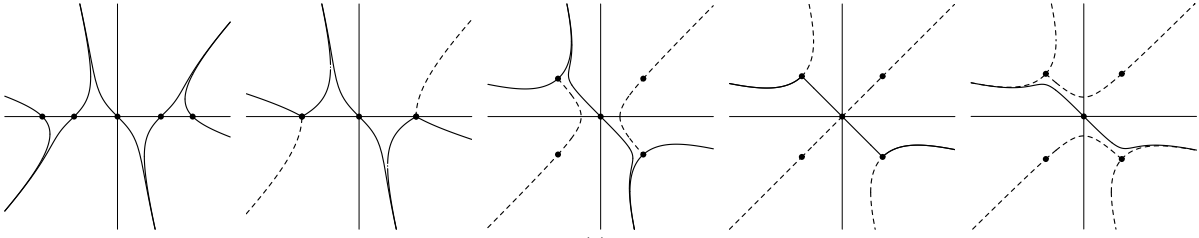}
\fi
\fi
\caption{Sequence of steepest--descent contours
for $\Phi^{(2)}(q,0)$ displaying a
tangent bifurcation and a Stokes transition.
Again there is a connection of the ghost satellite
and the central orbit as the condition for a Stokes transition is fulfilled.
Dashed lines indicate steepest--descent paths that are not needed to
connect the integration boundaries. A trick can be played with these
pictures to envision the situation for the tangent bifurcation at negative
$I$: The plots are rotated by 90 degrees (which corresponds to
inverting the sign of $a$),
and the contour is picked that originally connects $\pm i\infty$ (see
also Fig.\ \protect\ref{fig:p2cnt2}).
}
\label{fig:pver21cont}
\end{figure}

For even $m$ the problem is mapped on the case $m=2$.
We already determined the locations (\ref{eq:m2pos})
of the orbits for the
Hamiltonian normal form $H^{(2)}$. By convention,
coefficients changed sign in the
definition of $\Phi^{(2)}$, but the coordinates are not affected by this.
The satellites have the actions
\begin{equation}
S_\pm=S_0+\frac 13 \frac a b \varepsilon -\frac 2 {27} \frac {a^3}{b^2}
\pm 2b \left(\frac 19 \frac {a^2}{b^2}-\frac 13 \frac \varepsilon b
\right)^{3/2}
\;.
\end{equation}
From the condition $\mbox{Re}\,S_\pm=S_0$
one finds a Stokes transition of complex ghosts for 
$a=0$ if $\varepsilon b>0$. (No transition is encountered at
$\varepsilon=\frac 2 9 \frac {a^2}b$ or for $a=0$, but $\varepsilon b<0$,
since the radicand is positive then).
As for $m=1,3$ there is only one
scale--invariant parameter combination
$\varepsilon b/a^2$, but for each value there are now two variants of the
contour depending on $\mbox{sign}\,(ab)$.
The Stokes transition at $a=0$
involves a ghost with a nonvanishing imaginary part of the action.
Figure \ref{fig:pver21cont}
shows how the integration contour changes in the complex $q'$--plane
due to the tangent bifurcation and the
Stokes transition.
The plots proof the
existence of a path that connects both sheets at constant $\mbox{Re}\,\Phi$.

\begin{figure}
\ifprep
\epsfxsize5cm
\epsfbox{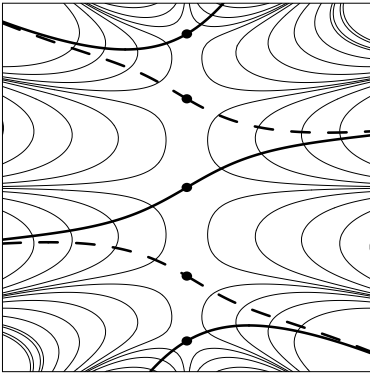}
\fi
\caption{No Stokes transition happens for ghosts
with real $I<0$ in the representative
case $m=2$. The steepest--descent
contours from the ghost
and the central orbit (full lines) are separated by the (dashed)
contour of another ghost at real, but negative $I$. Only the contour
that visits the central orbit is needed to connect the integration
boundaries.
}
\label{fig:p2cnt2}
\end{figure}

Different situations with $\mbox{Re}\,S_\pm=S_0$ appear,
however,
for $\varepsilon=\frac 1 4 \frac {a^2}b$.
No Stokes transition happens there
because the satellite involved is real for $ab<0$ and a
ghost with real action and real $I<0$ for $ab>0$. 
Figure \ref{fig:p2cnt2} confirms that indeed the contours are separated by
another `real ghost' (dashed contour).

For $m=5$
the phase function on the $p$--axis
is a
polynomial of degree five.
It appears on first sight
that the three
satellites could be arranged in such a way that the sheet of
the ghost $+$ is separated from the sheet of the central orbit $0$
by the remaining real satellite $1$.
One easily finds, however,
that the sheets are separated
for the particular phase function
only if
the real parts of the three roots $p_\pm$, $p_1$ of equation
(\ref{eq:m5fp})
all have the same sign,
which in turn can be
ruled out by a careful inspection of equation
(\ref{eq:m5fp}):
From Vieta's relations it follows that otherwise
the coefficient
of the linear term
$\sim [p_+ p_-+(p_++p_-)p_1]$
would not vanish.
Note that the order $I^{3/2}$ in $\Phi^{(5)}$
indeed results in a nonvanishing coefficient but
is considered, as usually, only as a perturbation
and does not alter the situation qualitatively.
The existence of a steepest--descent 
connection of the ghost and the central orbit at the Stokes transition
is then guaranteed also for $m=5$.
The transition takes place at the value of
$t\equiv 75 \varepsilon b^2 /(4 a^3)$
that is the solution of
\begin{equation}
640 + {\frac{1520\,t}{3}} +
{\frac{925\,{t^2}}{9}} + {t^3}=0
\end{equation}
close to $t=-97.6566$. To derive this equation
we first reduce the expression for
$\mbox{Re}\,S-S_0$ with help of the fixed--point equation
\begin{equation}
\label{eq:teq}
8 t+ 75 r^2(4+5r)=0
\end{equation}
for the scaled variable $r\sim p$
to the form $\mbox{Re\,}[300 r^2 + t(8-10r+75r^2)]=0$.
Introducing $r=(x+i y)$ here and in the
fixed--point equation and splitting the latter
into real and imaginary part, one can solve then for $y^2= 8 x/5 + 3
x^2$ and $t= 15 x (2+5x)^2$ and obtains the cubic equation 
$8+60x+125x^2+75x^3=0$. Its roots give three values
for $t$ which all solve (\ref{eq:teq}).
Only one of the roots, however, fulfills $y^2=8x/5+3>0$; this
is the one that corresponds to the approximate value of $t$ given above.

For odd $m\geq 7$
there are four different satellites on each line.
They are generically grouped in pairs that
lie on opposite sides of the
center, and 
the situation is similar to $m=2$ with slightly broken reflection
symmetry. Indeed it follows
from the derivation of the normal forms 
that the
odd terms originating from higher--order
perturbations are neglegible for small $\varepsilon$ and $a$;
the results for $m=2$ are then directly applicable to the
present case.

\section{Conclusions}
\label{sec:disc}

We studied
bifurcations of codimension two in Hamiltonian systems that are
either autonomous and have two degrees of freedom or periodic with
one degree of freedom.
The normal forms derived in section \ref{sec:nform} and discussed in
section
\ref{sec:discn} show that the
typical sequences of codimension one bifurcations in the neighbourhood 
(in parameter space) of the bifurcation of codimension two consists of
a period--$m$ bifurcation at a central orbit
followed by a tangent bifurcation
in which satellites become ghosts.

Additional generic scenarios are encountered in the presence of
symmetries
\cite{Rimmer:1983,Golubitsky:Stewart:1987,Aguiar:Malta:1987,Aguiar:Malta:1988%
,Ozorio:Aguiar:1990}.
Isochronous pitchfork bifurcations are
the most important addition of codimension one
in the case of time--reversal or reflection symmetries;
they will also
show up in the neighbourhood of codimension two bifurcations 
in these systems.

Only
a small number of the bifurcations of codimension one and 
two correspond to (special cases of) a
so--called elementary catastrophe due to Thom
(see e.\,g.\ \cite{Poston:Stewart:1978,Berry:Upstill:1980}).
These appear in many different contexts and
describe,
for instance, bifurcations of
codimension up to four in maps that are not restricted by area
preservation.
We use the usual names and symbols
and further denote each Hamiltonian bifurcation type by $(m_k)$,
where $m$ is the multiplicity and $k=1,2$ the codimension.
The fold $A_2$ corresponds to the tangent bifurcation $(1_1)$.
The cusp $A_3$ is $(1_2)$, and $(2_1)$ is a cusp with a reflection
symmetry.
$(2_2)$ is a butterfly $A_5$ with reflection symmetry.
The period--tripling bifurcation $(3_1)$ corresponds to
a version of the elliptic
umbilic $D_4^-$.  All other normal forms describe catastrophes that
would be of much higher codimension without area preservation.
Especially for the cases $m\geq 3$ 
one has to rely on higher--order perturbation
theory.
It implies that i) for a given
codimension the class of bifurcations in Hamiltonian systems
is considerably larger and
ii) although this can be circumvented by considering
a normal form of much higher codimension from ordinary
catastrophe theory, these normal forms have then again to be restricted:
Points that correspond to the trajectory of one and
the same orbit lie on the same height (energy or action).
The classical perturbation
theory takes care of this and in addition gives the right codimension.

Collective contributions to semiclassical traces
were derived that involve normal forms for a
phase function $\Phi$ and an amplitude function $\Psi$. The expressions
involve just as many coefficients as are determined by the actions and
stability properties of the bifurcating orbits, including a suppression
of certain unwanted ghosts for $m\geq 4$.
The expressions constitute uniform approximations: They are also valid
far away from the bifurcation and asymptotically take the form of a
sum of isolated contributions (\ref{eq:statphaseint}).
The uniform approximations display Stokes transitions
in which the ghost satellites interact once more with the central orbit and
leave the steepest--descent integration contour. 

The validity of the approximations given here
is limited if additional orbits become important or `unwanted' ghosts
become real; bifurcations of even
higher codimension are then to be studied. The basic steps would be
the same as in the present study: Derivation of Hamiltonian normal forms
that account for all bifurcating orbits; reduction of normal forms to
get rid of non--bifurcating orbits and to account for independent
stabilities and actions.
An important open question
is concerned with the complexity of periodic--orbit clusters
typically encountered in the quest of
resolving spectra 
when one approaches the semiclassical limit: This requires knowledge of
the dynamics up to the Heisenberg time $\sim 1/\hbar$ and involves a
competition of increasing resolution in phase space and proliferation of
periodic orbits. 

One might also be concerned about cascades, which are sequences of
bifurcations
of a certain orbit
of period $n$
at differing values of
$\mbox{tr}\,M^{(n)}=2\cos\omega$ (cf.\ section \ref{sec:bifcond}).
The
most prominent example is the basic building block
of period--doubling cascades:
an orbit of period $2n$ is born at an orbit of period $n$
in a period--doubling bifurcation
($\mbox{tr}\,M^{(2n)}=2$) and period--doubles itself  at
$\mbox{tr}\,M^{(2n)}=-2$. A huge variety of cascades exists, however,
since bifurcations happen whenever the stability angle $\omega$ is a
rational multiple of $2\pi$.
Bifurcations in a cascade cannot be encountered
simultaneously in parameter space,
since this would imply a singular change in the
linearized map. For that reason cascades cannot be regarded as
unfoldings of bifurcations of higher codimension. The bifurcations in
an unfolding show up simultaneously in a given iteration
of the map; the cascades involve bifurcations
that appear in distinct iterations.
One could study, for instance, those cascades
that arise from the iteration of the map generated by a
normal form, and
ask the question whether 
situations exist in which the orbits in the
cascade must be treated collectively; it would be indeed nice to see that one
can do without. An argument in favor of this expectation has been given in
\cite{Ozorio:Hannay:1987}.

\section*{Acknowledgments}

The author thanks F. Haake, C. Howls, J. Keating, D. Sadovski{\'\i},
and M. Sieber for helpful discussions.
Support by the Sonderforschungsbereich
`Unordnung und gro{\ss}e Fluktuationen' of the Deutsche Forschungsgemeinschaft
is gratefully acknowledged.

\begin{thebibliography}{10}

\bibitem{Gutzwiller:1971}
M.~C. Gutzwiller, J. Math. Phys. {\bf 12},  343  (1971).

\bibitem{Gutzwiller:1990}
M.~C. {Gutzwiller}, {\em Chaos in Classical and Quantum Mechanics} (Springer,
  New York, 1990).

\bibitem{Tabor:1983}
M. Tabor, Physica D {\bf 6},  195  (1983).

\bibitem{Junker:Leschke:1992}
G. Junker and H. Leschke, Physica D {\bf 56},  135  (1992).

\bibitem{Berry:Tabor:1976}
M.~V. Berry and M. Tabor, Proc. R. Soc. Lond. A {\bf 349},  101  (1976).

\bibitem{Berry:Tabor:1977}
M.~V. Berry and M. Tabor, J. Phys. A {\bf 10},  371  (1977).

\bibitem{Ozorio:Hannay:1987}
A.~M. {Ozorio de Almeida} and J.~H. {Hannay}, J. Phys. A {\bf 20},  5873
  (1987).

\bibitem{Ozorio:1988}
A.~M. {Ozorio de Almeida}, {\em {H}amiltonian Systems: Chaos and Quantization}
  (Cambridge University Press, Cambridge, 1988).

\bibitem{Kus:Haake:1993b}
M. Ku{\'s}, F. Haake, and D. Delande, Phys. Rev. Lett. {\bf 71},  2167  (1993).

\bibitem{Sieber:1996a}
M. {Sieber}, J. Phys. A {\bf 29},  4715  (1996).

\bibitem{Schomerus:Sieber:1997a}
H. Schomerus and M. Sieber, J. Phys. A {\bf 30},  4537  (1997).

\bibitem{Sieber:Schomerus:1997a}
M. Sieber and H. Schomerus, chao-dyn/9708013, accepted for publication in J. Phys. A (1997).

\bibitem{Meyer:1970}
K.~R. Meyer, Trans. Am. Math. Soc. {\bf 149},  95  (1970).

\bibitem{Bruno:1970}
A.~D. {Bruno}, Math. USSR Sbornik {\bf 12},  271  (1970).

\bibitem{Bruno:1972}
A.~D. {Bruno}, preprint Nr. 18, Inst. Prikl. Mat. Akad. Nauk SSSR, Moskau (in
  Russian) (1972).

\bibitem{Rimmer:1983}
R.~J. Rimmer, Memoirs of the AMS 272, American Mathematical Society,
  Providence, Rhode Island.

\bibitem{Golubitsky:Stewart:1987}
M. Golubitsky and I. Stewart, Physica D {\bf 24},  391  (1987).

\bibitem{Aguiar:Malta:1987}
M.~A.~M. {de Aguiar}, C.~P. Malta, M. Baranger, and K.~T.~R. Davies, Ann. Phys.
  (N. Y.) {\bf 180},  167  (1987).

\bibitem{Aguiar:Malta:1988}
M.~A.~M. {de Aguiar} and C.~P. Malta, Physica D {\bf 30},  413  (1988).

\bibitem{Ozorio:Aguiar:1990}
A.~M. {Ozorio de Almeida} and M.~A.~M. {de Aguiar}, Physica D {\bf 41},  391
  (1990).

\bibitem{Sadovskii:Shaw:1995}
D.~A. Sadovski{\'\i}, J.~A. Shaw, and J.~B. Delos, Phys. Rev. Lett. {\bf 75},
  2120  (1995).

\bibitem{Sadovskii:Delos:1996}
D.~A. Sadovski{\'\i} and J.~B. Delos, Phys. Rev. E {\bf 54},  2033  (1996).

\bibitem{Poincare:1957}
H. Poincar{\'e}, {\em New Methods of Celestial Mechanics} (Dover, New York,
  1957), Vol.~III.

\bibitem{Birkhoff:1927}
G.~D. Birkhoff, {\em Dynamical Systems} (American Mathematical Society, New
  York, 1927).

\bibitem{Deprit:1969}
A. Deprit, Celest. Mech. {\bf 1},  12  (1969).

\bibitem{Arnold:1988}
V.~I. {Arnol'd}, {\em Geometrical Methods of the Theory of Ordinary
  Differential Equations}, Vol.~250 of {\em Series of Comprehensive Studies in
  Mathematics} (Springer, New York, 1988).

\bibitem{Schomerus:1997a}
H. Schomerus, Europhys. Lett. {\bf 38},  423  (1997).

\bibitem{Schomerus:Haake:1997}
H. Schomerus and F. Haake, Phys. Rev. Lett. {\bf 79},  1022  (1997); a more detailed exposition is
in preparation.

\bibitem{Main:Wunner:1997}
J. Main and G. Wunner, Phys. Rev. A {\bf 55},  1743  (1997).

\bibitem{Berry:1989}
M.~V. Berry, Proc. R. Soc. Lond. A {\bf 422},  7  (1989).

\bibitem{Boasman:Keating:1995}
P.~A. Boasman and J.~P. Keating, Proc. R. Soc. Lond. A {\bf 449},  629  (1995).

\bibitem{Meyer:Hall:1992}
K.~R. Meyer and G.~R. Hall, {\em Introduction to {H}amiltonian Dynamical
  Systems and the $N$-Body Problem}, Vol.~90 of {\em Applied Mathematical
  Sciences} (Springer, New York, 1992).

\bibitem{Mao:Delos:1992}
J.-M. Mao and J.~B. Delos, Phys. Rev. A {\bf 45},  1746  (1992).

\bibitem{Tomsovic:Grinberg:1995}
S. Tomsovic, M. Grinberg, and D. Ullmo, Phys. Rev. Lett. {\bf 75},  4346
  (1995).

\bibitem{Ullmo:Grinberg:1996}
D. Ullmo, M. Grinberg, and S. Tomsovic, Phys. Rev. E {\bf 54},  136  (1996).

\bibitem{Wright:1980}
F.~J. Wright, J. Phys. A {\bf 13},  2913  (1980).

\bibitem{Poston:Stewart:1978}
T. Poston and I.~N. Stewart, {\em Catastrophe Theory and its Applications}
  (Pitman, London, 1978).

\bibitem{Berry:Upstill:1980}
M.~V. Berry and C. Upstill,  in {\em Progress in Optics}, edited by E. Wolf
  (North-Holland, Amsterdam, 1980), Vol.~VIII, pp.\ 257--346.

\end{thebibliography}

\ifprep
\onecolumn
\fi

\ifprepgood
\else
\vspace{-.5cm}
\fi

\end{document}